\documentclass[11pt]{article}
\pdfoutput=1
\usepackage[]{inputenc}
\usepackage[T1]{fontenc}
\usepackage{hyperref}
\usepackage{color}
\usepackage{fullpage}

\usepackage{amsmath}
\usepackage{amssymb}

\newcommand{\GG}{\ensuremath{\mathcal{G}}}
\newcommand{\LL}{\ensuremath{\mathcal{L}}}

\def \be {\begin{equation}}
\def \ee {\end{equation}}
\def \bea {\begin{eqnarray}}
\def \eea {\end{eqnarray}}

\title{On the local well-posedness of Lovelock and Horndeski theories}
\author{Giuseppe Papallo\footnote{\href{mailto:g.papallo@damtp.cam.ac.uk}{\nolinkurl{g.papallo@damtp.cam.ac.uk}}}\, and Harvey S. Reall\footnote{\href{mailto:h.s.reall@damtp.cam.ac.uk}{\nolinkurl{h.s.reall@damtp.cam.ac.uk}}} \\
\\
\it DAMTP, Centre for Mathematical Sciences, University of Cambridge, \\
\it Wilberforce Road, Cambridge CB3 0WA, UK}

\begin{document}
\maketitle

\begin{abstract}
We investigate local well-posedness of the initial value problem for Lovelock and Horndeski theories of gravity. A necessary condition for local well-posedness is strong hyperbolicity of the equations of motion. Even weak hyperbolicity can fail for strong fields so we restrict to weak fields. The Einstein equation is known to be strongly hyperbolic in harmonic gauge so we study Lovelock theories in harmonic gauge. We show that the equation of motion is always weakly hyperbolic for weak fields but, in a generic weak-field background, it is not strongly hyperbolic. For Horndeski theories, we prove that, for weak fields, the equation of motion is always weakly hyperbolic in any generalized harmonic gauge. For some Horndeski theories there exists a generalized harmonic gauge for which the equation of motion is strongly hyperbolic in a weak-field background. This includes ``k-essence'' like theories. However, for more general Horndeski theories, there is no generalized harmonic gauge for which the equation of motion is strongly hyperbolic in a generic weak-field background. Our results show that the standard method used to establish local well-posedness of the Einstein equation does not extend to Lovelock or general Horndeski theories. This raises the possibility that these theories may not admit a well-posed initial value problem even for weak fields. 
\end{abstract}

\section{Introduction}

Lovelock theories of gravity are the most general diffeomorphism-covariant theories involving a metric tensor with second order equations of motion \cite{Lovelock1971}. In four dimensions, the equation of motion of such a theory reduces to the Einstein equation. But in higher dimensions extra terms are present and these can change significantly the properties of the equation. For example, it is well-known that in these theories, gravity does not travel at the speed of light, instead the speed depends on the curvature of spacetime \cite{Aragone:1987jm,Choquet-Bruhat1988}. 

Horndeski theories are the most general four-dimensional diffeomorphism-covariant theories involving a metric tensor and a scalar field, with second order equations of motion \cite{Horndeski1974}. Some of these theories can be obtained from  Lovelock theories by dimensional reduction. 

Although Lovelock and Horndeski theories have been discussed extensively, the issue of their mathematical consistency has not received much attention. A minimal consistency requirement of a classical theory is that the initial value problem should be locally well-posed. This means that, given suitable initial data, there should exist a unique solution of the equation of motion arising from the data, and this solution should depend continuously on the data. ``Local'' here means that the solution is only required to exist for some non-zero time, no matter how small.

For {\it analytic} initial data, local existence and uniqueness of solutions can be established straightforwardly -- this was done for Lovelock theories in Ref.~\cite{Choquet-Bruhat1988}. However, this does not establish continuous dependence of the solution on the data. Furthermore, the restriction to analytic data is unphysical because it implies that the solution is determined everywhere by its behaviour at a single point. One cannot discuss causality if one restricts to such data. 

Consider the problem of {\it nonlinear} perturbations of some ``background'' solution of a Lovelock or Horndeski theory. For the nonlinear initial value problem to be well-posed, it is necessary that the initial value problem for {\it linearized} perturbations should also be locally well-posed, not just around the background solution but around any solution in a neighbourhood of this background solution.

For the linearized initial value problem to be well-posed, the equation of motion should be {\it hyperbolic}, i.e., have the character of a wave equation. Two notions of hyperbolicity can be distinguished \cite{Kreiss1989,Sarbach2012}. Roughly speaking, an equation is {\it weakly} hyperbolic if it never admits solutions which grow exponentially in time, with the exponent proportional to the magnitude of a spatial wavevector, i.e., growth which is arbitrarily fast at arbitrarily short distances. An equation is {\it strongly} hyperbolic if an appropriate norm of the solution at time $t$ can be bounded by the initial value of the same norm multiplied by a function of time which is independent of the initial data. Such a bound is an example of an {\it energy estimate}. Obtaining such an estimate is the standard way of proving local well-posedness. Note that strong hyperbolicity implies weak hyperbolicity.  

For a diffeomorphism-covariant theory, the gauge freedom implies that the equation of motion will not be hyperbolic unless one imposes an appropriate gauge condition. For the Einstein equation, the simplest choice is harmonic gauge, which ensures that the equation is strongly hyperbolic, and one can establish local well-posedness \cite{Choquet-Bruhat1952}. Other approaches to the Einstein equation, such as the ADM formulation \cite{Arnowitt:1962hi}, give equations which are weakly but not strongly hyperbolic \cite{Nagy2004}. This implies that they cannot be used to establish local well-posedness.\footnote{However there exist strongly hyperbolic modifications of these equations which can be used to establish local well-posedness \cite{Rodnianski2014a}.} It also means that they are unsuitable for solving the Einstein equation numerically on a computer. For numerical applications, strong hyperbolicity is regarded as essential. The first successful binary black hole simulations \cite{Pretorius:2005gq,Campanelli:2005dd,Baker:2005vv} employed numerical codes based either on harmonic gauge \cite{Pretorius:2005gq} or the BSSN formalism \cite{Nakamura:1987zz,Shibata:1995we,Baumgarte:1998te}. The latter is a modification of the ADM formalism that can be shown to be strongly hyperbolic \cite{Gundlach:2006tw}.

We will start by discussing weak hyperbolicity of (linearized) Lovelock and Horndeski theories. The results of previous work shows that weak hyperbolicity can fail if the background fields become too large. It was shown in Ref. \cite{Reall2014} that weak hyperbolicity fails (in any gauge) for linear perturbations of ``small'' black hole solutions of Lovelock theories. Here ``small'' refers to the scale set by the dimensionful coupling constants of such a theory. More generally, one expects that weak hyperbolicity will fail in a large class of backgrounds with large curvature. In Horndeski theories, it has been shown that cosmological solutions can suffer from ``ghost and gradient instabilities'' when the fields become large \cite{DeFelice2010,DeFelice2011,Kobayashi2011}. As we will explain below, these ``instabilities'' are not dynamical instabilities but instead indicate a failure of weak hyperbolicity in such backgrounds. These examples show that, for both Lovelock and Horndeski theories, the equation of motion is {\it not} always weakly hyperbolic. Hence for general initial data one cannot expect local well-posedness. However, one might hope that if one restricts the initial data so that the equation of motion {\it is} weakly hyperbolic then the initial value problem will be locally well-posed. In particular, one might expect that a failure of weak hyperbolicity would occur only for large background fields so that if we restrict to studying backgrounds involving only {\it weak} fields then there will be no problem. In this paper we will investigate whether or not this is true.

For weak fields, the equation of motion of a Lovelock or Horndeski theory appears to be a small perturbation of the Einstein equation and therefore one might guess that the equation of motion will be hyperbolic. However, this is not obvious because the perturbation to the equation of motion changes the two-derivative terms. We can illustrate this point with an example. In 2d Minkowski spacetime consider the equations
\be
\label{system}
 \partial^2 \phi = k \epsilon \partial_0 \partial_1 \psi \qquad \partial^2 \psi = - k \epsilon \partial_0 \partial_1 \phi.
\ee
View this system as analogous to the equations governing linear perturbations around a weak field background solution of Lovelock or Horndeski theory. Here $k$ is to be regarded as analogous to a coupling constant of the theory, with $k=0$ analogous to the Einstein equation. The parameter $\epsilon$ corresponds to the strength of the background fields, with the Lorentz symmetry breaking on the RHS analogous to the Lorentz symmetry breaking arising from the non-trivial backgrounds fields. For $k=0$ we have a hyperbolic system. However, when $k \ne 0$, for any $\epsilon \ne 0$ it is easy to check that the above system of equations is {\it elliptic}. This example demonstrates that a small perturbation to the highest derivative terms in an equation of motion can completely change the character of the equation.\footnote{
There is, however, an important difference between the system (\ref{system}) and a Lovelock or Horndeski theory, which is that (\ref{system}) is not obtained from an action principle.}

We will show that the above problem does {\it not} occur for Lovelock or Horndeski theories. We will prove that these theories are weakly hyperbolic in any weak field background. More precisely, we will prove that the linearized equation of motion is weakly hyperbolic in harmonic gauge (Lovelock) or a generalized harmonic gauge (Horndeski) whenever the background fields are sufficiently weak.

Our most important results concern strong hyperbolicity of Lovelock and Horndeski theories. As discussed above, strong hyperbolicity is needed in order to establish local well-posedness of the initial value problem, and in numerical applications. However, we will prove that, for Lovelock theories, in harmonic gauge, the linearized equation of motion is {\it not} strongly hyperbolic in a generic weakly curved background. The word ``generic'' is important here: there certainly exist particular backgrounds for which the linearized equation of motion {\it is} strongly hyperbolic (e.g. Minkowski spacetime \cite{Boulware1985}) so the equation of motion for linear perturbations around such backgrounds is locally well-posed. However, such backgrounds are non-generic e.g. they always have symmetries. In order to have any hope of establishing local well-posedness for the {\it nonlinear} theory for weak fields, one would need strong hyperbolicity for {\it any} weakly curved background. This is not the case, at least not in harmonic gauge. Hence the most straightforward approach to establishing local well-posedness for Lovelock theories does not work.\footnote{Note that the recent discussion of local well-posedness in Ref. \cite{Willison2014} simply {\it assumes} that the harmonic gauge equation of motion is suitably hyperbolic. Our result shows that this assumption is incorrect.} In the final section of this paper we will discuss whether any alternative method could work. 

For a particular class of Horndeski theories, we will prove that there exists a generalized harmonic gauge for which the linearized equation of motion {\it is} strongly hyperbolic for arbitrary weak background fields. This class of theories involves no coupling between derivatives of the scalar field and curvature tensors in the action. This class includes various models of interest e.g. ``k-essence'' theories or scalar-tensor theories such as Brans-Dicke theory \cite{Brans1961}. However, for more general Horndeski theories, we find that the situation is analogous to the Lovelock case: there exists no generalized harmonic gauge for which the linearized theory is strongly hyperbolic in a generic weak field background. 

This result can be strengthened considerably as follows. Consider a Horndeski theory for which there exists a generalized harmonic gauge such that the linearized equation of motion is strongly hyperbolic in a generic weak field background. We can now ask: does this extend to the nonlinear theory? In particular, does there exist a generalized harmonic gauge for the nonlinear theory such that the nonlinear equation of motion is strongly hyperbolic in a generic weak-field background? For this to be the case, the generalized harmonic gauge condition for the nonlinear theory must, upon linearization, reduce to the generalized harmonic gauge condition for the linearized theory. However, this implies that the source function appearing in the gauge condition of the linearized theory must satisfy a certain integrability condition. This condition is not satisfied in general. Using this condition we find that the class of Horndeski theories for which there exists a generalized harmonic gauge for which the nonlinear theory is strongly hyperbolic in a generic weak-field background is simply the class of ``k-essence'' type theories coupled to Einstein gravity. See the end of section \ref{sec:horndeski} for a precise statement. 

This paper is organized as follows: in Section~\ref{sec:hyperbolicity} we define the notions of weak and strong hyperbolicity and discuss the relevant background material. In Section~\ref{sec:lovelock} we discuss Lovelock theories. We present a proof of weak hyperbolicity of harmonic gauge Lovelock theories in weak curvature backgrounds and show that, generically, strong hyperbolicity does not hold. We then present some examples in which weak hyperbolicity is violated dynamically. These are ``collapsing universe'' solutions which start off with small curvature, but develop large curvature over time. In Section \ref{sec:horndeski} we discuss Horndeski theories. We show that, in a generalized harmonic gauge, Horndeski theories are weakly hyperbolic in weak-field backgrounds. We then show that, while a subclass of Horndeski theories is strongly hyperbolic in a particular generalized harmonic gauge, more general Horndeski theories are not. Section \ref{sec:discuss} discusses the implications of our results. 

We adopt the notation that Latin indices $a,b,c,\ldots$ are abstract indices denoting tensor equations valid in any basis. Greek indices $\mu,\nu,\ldots$ refer to a particular basis e.g. a coordinate basis. 
\section{Hyperbolicity}
\label{sec:hyperbolicity}

In this section we will review briefly the notions of weak and strong hyperbolicity. Our discussion is based on Refs.~\cite{Kreiss1989} and \cite{Sarbach2012}. We will start with first order systems of linear equations and then discuss second order systems. 

\subsection{First order equation}

\label{sec:firstorder}

In $d$ spacetime dimensions with coordinates $(t,x^i)$, consider a first order linear partial differential equation for a $N$-dimensional vector $u$:
\be
\label{firstorder}
A u_t + P^i \partial_i u + C u = 0
\ee
where $A$, $P^i$ and $C$ are real constant $N \times N$ matrices. We assume that $A$ is invertible. Taking a spatial Fourier transform gives
\be
  \tilde{u}_t - i {\cal M}(\xi_i) \tilde{u}  = 0
\ee
where
\be
 {\cal M}(\xi_i) = A^{-1} \left( - P^i \xi_i +i C \right).
\ee
This has solution
\be
 \tilde{u}(t,\xi_i) = \exp(  i {\cal M}(\xi_i) t) \tilde{u}(0,\xi_i) 
\ee
and hence we have the formal solution 
\be
\label{usol}
 u(t,x) =\frac{1}{(2\pi)^{d-1}}  \int d^{d-1} \xi \exp(- i \xi_i x^i)  \exp(  i {\cal M}(\xi_i) t) \hat{u}(0,\xi_i) .
\ee
The problem with this expression is that it may not converge as $|\xi| \rightarrow \infty$ without imposing unreasonable conditions on the initial data. Here we have defined
\be
 |\xi| = \sqrt{\xi_i\xi_i}.
\ee
To ensure convergence we need the matrix ${\cal M}(\xi_i)$ to satisfy certain conditions. Convergence is guaranteed if ${\cal M}(\xi_i)$ satisfies, for all $\xi_i$, and all $t>0$,
\be
\label{Rcond}
 ||  \exp(  i {\cal M}(\xi_i) t) || \le f(t)
\ee
for some continuous function $f(t)$ independent of $\xi_i$. This condition implies that the integral converges and the resulting solution satisfies
\be
\label{L2est}
|| u ||(t) \le f(t) ||u||(0)
\ee
where $||\ldots ||$ denotes the spatial $L^2$ norm. Using this one can prove that the initial value problem is locally well-posed. So we need to determine whether or not the condition (\ref{Rcond}) is satisfied. 
 
The convergence of (\ref{usol}) is a high frequency question so we let $t = t'/|\xi|$ and take $|\xi| \rightarrow \infty$ at fixed $t'$. Equation~\eqref{Rcond} becomes
\be
\label{Rcond2}
 || \exp (i M (\hat{\xi_i}) t')|| \le k,
\ee
where $k = f(0)$,
\be
\hat{\xi}_i = \frac{\xi_i}{|\xi|},
\ee
and
\be
M(\xi_i)= -A^{-1} P^i \xi_i
\ee
is the ``principal part'' of ${\cal M}$. Consider an eigenvector $v$ of $M(\hat{\xi}_i)$ with eigenvalue $\lambda = \lambda_1  + i \lambda_2$. We have
\be
  \exp(  i M(\hat{\xi}_i) t') v = e^{i \lambda_1 t' } e^{-\lambda_2  t'} v.
\ee
This is consistent with (\ref{Rcond2}) only if $ \lambda_2 \ge 0$ for all $\hat{\xi}$. Now $M(\hat{\xi}_i)$ is a real matrix hence if $\lambda$ is an eigenvalue then so is $\bar{\lambda}$ so consistency with (\ref{Rcond2}) requires $\pm \lambda_2 \ge 0$ i.e. $\lambda_2=0$. We deduce that (\ref{Rcond2}) implies that all eigenvalues of $M(\hat{\xi}_i)$ are real. This motivates the definition of {\it weak hyperbolicity}: equation (\ref{firstorder}) is {\it weakly hyperbolic} if, and only if, all eigenvalues of $M(\xi_i)$ are real for any real $\xi_i$ with $\xi_i \xi_i = 1$. 

A failure of weak hyperbolicity would be a disaster for the initial value problem because the integrand in (\ref{usol}) would grow exponentially with $|\xi|$ at large $|\xi|$ so convergence would require highly fine-tuned initial data. 

The matrix $M(\hat{\xi}_i)$ can be brought to Jordan normal form by a similarity transformation
\be
 M(\hat{\xi}_i) = S(\hat{\xi}_i) J(\hat{\xi}_i) S(\hat{\xi}_i)^{-1}
\ee
so
\be
 \exp (i M (\hat{\xi}_i) t') = S(\hat{\xi}_i) \exp (i J (\hat{\xi}_i) t')S(\hat{\xi}_i)^{-1}.
\ee
Assume that $M$ is not diagonalizable, i.e., $J$ contains a $n \times n$ Jordan block, $n \ge 2$, associated to an eigenvalue $\lambda$ of $M(\hat{\xi})$. Then the RHS exhibits polynomial growth with $t'$. For example consider the case of a $2\times 2$ block $J_2$ with eigenvalue $\lambda$:
\be
J_2 = \left( \begin{array}{ll} \lambda & 1 \\ 0 & \lambda \end{array} \right) \qquad \Rightarrow \qquad  \exp (i J_2 t') = e^{i\lambda t'}  \left( \begin{array}{ll} 1 & it' \\ 0 & 1 \end{array} \right) .
\ee
If the equation is weakly hyperbolic then $\lambda$ is real so there is no exponential growth in $t'$. But the presence here of the term linear in $t'$ implies that equation (\ref{Rcond2}) is not satisfied. More generally, an $n \times n$ Jordan block would lead to terms involving $t'^p$ for $p \le n$. Using $t' = |\xi|t$ this gives terms proportional to $|\xi|^p$ in the integral of (\ref{usol}). The presence of such terms implies that it is not possible to obtain a bound of the form (\ref{L2est}). The best one could hope for is that it is possible to modify the RHS to include sufficiently many spatial derivatives of $u$. Whether or not this is possible depends on the form of the zero derivative term $Cu$ in the equation of motion \cite{Kreiss1989}.\footnote{
There are examples of weakly hyperbolic systems for which $|| \exp(  i {\cal M}(\xi_i) t) ||$ grows as $\exp{(c \sqrt{|\xi|} t)}$ for some constant $c>0$ \cite{Kreiss1989}, in which case one could not even obtain a bound of this weaker type.} But even if this is possible, the ``loss of derivatives'' in (\ref{L2est}) would be worrying if we are considering an equation obtained by linearizing some nonlinear equation. This is because the loss of derivatives would be a serious obstruction to establishing local well-posedness for the {\it nonlinear} equation. 

To avoid this problem, $M(\hat{\xi})$ must be diagonalizable, i.e., there exists a matrix $S(\hat{\xi}_i)$ such that $M = SDS^{-1}$ where $D(\hat{\xi}_i)$ is diagonal. Defining a positive definite Hermitian matrix $K(\hat{\xi}_i) = (S^{-1})^\dagger S^{-1}$  we then have
\be
\label{Hcond}
 K(\hat{\xi}_i) M(\hat{\xi}_i) K(\hat{\xi}_i)^{-1} = M(\hat{\xi}_i)^\dagger.
\ee   
This motivates the definition of strong hyperbolicity. With constant coefficients, equation (\ref{firstorder}) is {\it strongly hyperbolic} if, and only if, there exists a positive definite Hermitian matrix $K(\hat{\xi}_i)$ depending smoothly on $\hat{\xi}_i$ such that (\ref{Hcond}) holds. 

Note that (\ref{Hcond}) states that $M(\hat{\xi}_i)$ is Hermitian w.r.t. $K(\hat{\xi}_i)$. This implies that $M(\hat{\xi}_i)$ is diagonalizable with real eigenvalues. Using $K$ one can define an inner product between solutions, and the corresponding norm can be shown to satisfy an inequality of the form (\ref{L2est}). This is called the {\it energy estimate}. Using this one can prove that the initial value problem is locally well-posed independently of the form of the zero derivative term $Cu$ in (\ref{firstorder}) \cite{Kreiss1989}. 

So far the discussion has considered a first order linear PDE with constant coefficients. We can now discuss the case of non-constant coefficients. Let the matrices $A$, $P^i$ and $C$ in (\ref{firstorder}) depend smoothly on time and space, with $A$ still invertible. 
At some point $(t_{0},x_0^i)$ we define the {\it frozen coefficients} equation by fixing $A$, $P^i$ and $C$ to their values at $(t_0,x^i_0)$. It is believed that a necessary condition for local well-posedness of the varying coefficients equation near $(t_0,x^i_0)$ is that the frozen coefficients equation should be locally well-posed. For this to be the case, the frozen coefficients equation must satisfy the above definitions of weak and strong hyperbolicity. For the varying coefficients equation to be locally well-posed, we need these definitions to be satisfied for all $(t_0,x^i_0)$. This motivates extending the definitions of hyperbolicity to equations with non-constant coefficients in the obvious way: we simply allow $M(t,x,\xi_i)$ and $K(t,x,\xi_i)$ to depend smoothly on $(t,x)$ as well as on $\xi$ \cite{Kreiss1989,Sarbach2012}:

\medskip

\noindent{\bf Definition.} {\it Equation (\ref{firstorder}) is {\rm weakly hyperbolic} if, and only if, all eigenvalues of $M(t,x,\xi_i)$ are real for any real $\xi_i$ with $\xi_i \xi_i = 1$.}

\medskip

\noindent {\it Equation (\ref{firstorder}) is {\rm strongly hyperbolic} if, and only if, there exists a positive definite Hermitian matrix $K(t,x,\hat{\xi}_i)$ depending smoothly on $t,x,\hat{\xi}_i$ such that 
\be
 K(t,x,\hat{\xi}_i) M(t,x,\hat{\xi}_i) K(t,x,\hat{\xi}_i)^{-1} = M(t,x,\hat{\xi}_i)^\dagger
\ee   
and a constant $C>0$ such that $C^{-1} I \le K(t,x,\hat{\xi}_i) \le C I$ for all $t,x,\hat{\xi}_i$. }

\medskip

The latter technical condition is required to obtain an energy estimate -- it ensures that $K$ does not behave badly for large $t,x$ e.g. it does not become degenerate or blow up asymptotically. 

\subsection{Second order equations}

\label{sec:secondorder}

Our treatment of second order equations is based on \cite{Sarbach2012}. Consider a second order linear equation for an $N$-dimensional vector $u$ in $d$ coordinates $x^\mu = (t,x^i)$:
\be
\label{secondorder}
P^{\mu\nu} \partial_\mu \partial_\nu u + Q^\mu \partial_\mu u + R u = 0
\ee
where $P^{\mu\nu}=P^{\nu\mu}$, $Q^\mu$ and $R$ are $N \times N$ real matrices. For a covector $\xi$, the {\it principal symbol} is the matrix
\be
 P(\xi) \equiv P^{\mu\nu} \xi_\mu \xi_\nu.
\ee
As above, we start by considering the constant coefficients case. Take a spatial Fourier transform to obtain
\be
A \tilde{u}_{tt} + i \left( B(\xi_i) + i Q^0 \right) \tilde{u}_t - \left( C(\xi_i) + i Q^i \xi_i + R \right) \tilde{u}=0
\ee
where
\be
\label{ABCdef}
A = P^{00} \qquad B(\xi_i) = 2\xi_i P^{0i} \qquad C(\xi_i) = P^{ij} \xi_i \xi_j 
\ee 
and we assume that $A$ is invertible, i.e., surfaces of constant $t$ are non-characteristic. We write this equation in first order form by defining a column vector $\tilde{w}$ by\footnote{Here we slightly modify the approach of \cite{Sarbach2012} to avoid singularities at $|\xi|=0$.} 
\be
\tilde{w}^T = (\sqrt{1+|\xi|^2} \tilde{u},-i \tilde{u}_t)
\ee
where, as above,
\be
 |\xi| = \sqrt{\xi_i \xi_i},
\ee
giving the equation
\be
 \tilde{w}_t = i {\cal M}(\xi_i) \tilde{w}
\ee
where we define the $2N \times 2N$ matrix
\be
 {\cal M}(\xi_i) =  \begin{pmatrix} 0 & (1+|\xi|^2)^{1/2} I \\ -(1+|\xi|^2)^{-1/2} A^{-1} \left( C(\xi_i) + i Q^i \xi_i + R \right) & - A^{-1} \left( B(\xi_i)+ iQ^0 \right) \end{pmatrix} .
\ee
Note that the $L^2$ norm of $\hat{w}$ is a measure of the energy of the field $u$: it is quadratic in $u$ and its first derivatives. To prove local well-posedness requires that this norm obeys an energy estimate of the form
\be
 ||\tilde{w}||(t) \le f(t) ||\tilde{w}||(0)
\ee
for some continuous function  $f(t)$ independent of $\xi_i$ and $\tilde{w}$. 

The solution of the first order equation is
\be
 \tilde{w}(t,\xi_i) = \exp(  i {\cal M}(\xi_i) t) \tilde{w}(0,\xi_i) 
\ee
so for the energy estimate to hold for any initial data we need
\be
 || \exp(  i {\cal M}(\xi_i) t) || \le f(t).
\ee
If we define $t = t'/|\xi|$ at take $|\xi| \rightarrow \infty$ at fixed $t'$ then this implies
\be
\label{Mineq}
 || \exp(  i M(\hat{\xi_i}) t') || \le k
\ee
where $k=f(0)$, $\hat{\xi}_i =\xi_i/|\xi|$ and 
\be
\label{Mdef}
 M(\xi_i) =  \begin{pmatrix} 0 & I \\ - A^{-1}  C(\xi_i) & - A^{-1} B(\xi_i) \end{pmatrix} .
\ee
We can now repeat the argument we used for a first order system: if $M(\hat{\xi}_i)$ had a complex eigenvalue then we could violate (\ref{Mineq}). Hence we define weak hyperbolicity as the condition that all eigenvalues of $M(\hat{\xi_i})$ are real. 

Let $\xi_0$ be an eigenvalue of $M(\xi_i)$ with eigenvector $(t,t')^T$. Writing out the eigenvalue equation gives $t' = \xi_0 t$ and 
\be
 \left( A \xi_0^2 + B (\xi_i) \xi_0 + C(\xi_i) \right) t=0.
\ee
This is a {\it quadratic eigenvalue problem} with eigenvector $t$. In terms of the principal symbol it is simply
\be
 P(\xi) t =0 
\ee
where $\xi_\mu = (\xi_0,\xi_i)$. This equation states that the covector $\xi$ is {\it characteristic}. Hence (\ref{secondorder}) is {\it weakly hyperbolic} if, for any real $\xi_i \ne 0$, a characteristic covector $(\xi_0,\xi_i)$ has real $\xi_0$. 

As for first order systems, if the Jordan normal form of $M$ involves non-trivial blocks then equation (\ref{Mineq}) cannot hold. So we define strong hyperbolicity just as we did above: 
equation (\ref{secondorder}) is {\it strongly hyperbolic} if, and only if, there exists a positive definite Hermitian matrix $K(\hat{\xi}_i)$ depending smoothly on $\hat{\xi}_i$ such that $M(\hat{\xi}_i)$ is Hermitian w.r.t. $K$, i.e., satisfies (\ref{Hcond}). This implies that $M(\xi_i)=|\xi|^2 M(\hat{\xi}_i)$ is diagonalizable with real eigenvalues.

Finally we consider the equation (\ref{secondorder}) with coefficients $P^{\mu\nu}$, $Q^\mu$ and $R$ now depending on $(t,x^i)$. We define $M(t,x,\xi_i)$ using (\ref{Mdef}). As for first order systems, it is believed that local well-posedness implies local well-posedness for the equation with frozen coefficients. Hence we define weak and strong hyperbolicity just as for first order systems:

\medskip

\noindent{\bf Definition.} {\it Equation (\ref{secondorder}) is {\rm weakly hyperbolic} if, and only if, all eigenvalues of $M(t,x,\xi_i)$ are real for any real $\xi_i$ with $\xi_i \xi_i =1$. Equivalently, if $(\xi_0,\xi_i)$ is characteristic and $\xi_i \ne 0$ is real then $\xi_0$ is real. }

\medskip

\noindent {\it Equation (\ref{secondorder}) is {\rm strongly hyperbolic} if, and only if, there exists a positive definite Hermitian matrix $K(t,x,\hat{\xi}_i)$ depending smoothly on $t,x,\hat{\xi}_i$ such that 
\be
\label{Hcond2}
 K(t,x,\hat{\xi}_i) M(t,x,\hat{\xi}_i) K(t,x,\hat{\xi}_i)^{-1} = M(t,x,\hat{\xi}_i)^\dagger
\ee   
and a constant $C>0$ such that $C^{-1} I \le K(t,x,\hat{\xi}_i) \le C I$ for all $t,x,\hat{\xi}_i$. }

\medskip

In this paper we will mainly be interested in showing that certain equations are {\it not} strongly hyperbolic. We will do this by demonstrating that $M(t,x,\hat{\xi}_i)$ is not diagonalizable. Note that $M$ is determined by $P^{\mu\nu}$, i.e., by the principal symbol. So hyperbolicity depends only on the nature of the second derivative terms in the equation. Furthermore, to demonstrate that $M$ is not diagonalizable it is sufficient to work at a single point in spacetime. 

\section{Lovelock theories}

\label{sec:lovelock}

\subsection{Equation of motion in harmonic gauge}

In $d>4$ spacetime dimensions, the equation of motion of a Lovelock theory of gravity is
\be
A_{ab} = 8\pi T_{ab},
\ee
where $T_{ab}$ is the energy momentum tensor of matter and 
\be
\label{eq:lovelock_tensor}
 A^a{}_b =  G^a{}_b + \Lambda \delta^a_b + \sum_{p \ge 2} k_p \delta^{a c_1 \ldots c_{2p}}_{b d_1 \ldots d_{2p}} R_{c_1 c_2}{}^{d_1 d_2} \ldots R_{c_{2p-1} c_{2p}}{}^{d_{2p-1} d_{2p}} .
\ee
We have assumed that the coefficient of the Einstein term is non-zero and normalized it in the standard way. $k_p$ are constants and the antisymmetry ensures that the sum is finite ($2p+1 \le d$ in $d$ dimensions). We will be considering the case of vacuum solutions of this theory so we set $T_{ab}=0$ henceforth.

To investigate hyperbolicity we linearize around a background solution $g_{ab}$, i.e. the metric is $g_{ab} + h_{ab}$ and we linearize in $h_{ab}$, writing
\be
 A_{ab}[g+h] = A_{ab}[g] + A_{ab}^{(1)}[h] + \ldots
\ee
so that the linearized equation of motion is
\be
\label{linearized}
 A_{ab}^{(1)}[h] =0 .
\ee
For the Einstein equation (i.e. $k_p=0$), the resulting equation is strongly hyperbolic only if we impose a suitable gauge condition. For the {\it nonlinear} equation, one can choose {\it harmonic coordinates}: 
\be
 0 =  g^{\nu\rho} \nabla_\nu \nabla_\rho x^\mu = \frac{1}{\sqrt{-g}} \partial_\nu \left( \sqrt{-g} g^{\mu\nu} \right).
\ee
Upon linearization this reduces to the Lorenz gauge condition for the linearized metric perturbation:
\be
\label{harmonic}
 H_b \equiv \nabla^a h_{ab} - \frac{1}{2} \nabla_b h^a_a =0.
\ee
Actually, linearizing the harmonic gauge condition around a non-trivial background gives a {\it generalized} Lorenz gauge condition with a non-vanishing RHS. But this RHS does not depend on derivatives of $h_{ab}$ which implies that it does not affect the hyperbolicity analysis. Therefore we will just use the standard Lorenz gauge. 

Although most properly referred to as Lorenz gauge, henceforth we will refer to \eqref{harmonic} as harmonic gauge because it is inconvenient to use different words for the linear and nonlinear gauge conditions. Of course, it is well-known that the gauge condition \eqref{harmonic} can always be achieved by a suitable gauge transformation in the linearized theory. 

In harmonic gauge, the Einstein equation is strongly hyperbolic. We will investigate whether the same is true for Lovelock theory. We will do this by investigating hyperbolicity of the linearized theory. The harmonic gauge linearized equation of motion is 
\be
\label{harmoniceqofmotion}
 \tilde{A}_{ab}^{(1)}[h]=0,
\ee
where
\be
\label{Atdef}
 \tilde{A}_{ab}^{(1)}[h] \equiv A_{ab}^{(1)}[h] - \nabla_{(a} H_{b)} + \frac{1}{2} g_{ab} \nabla^c H_c
\ee
This is the equation of motion whose hyperbolicity we will study. 

A standard argument shows that the harmonic gauge condition is propagated by the harmonic gauge equation of motion \cite{Choquet-Bruhat1988}. The argument is based on the fact that the tensor $A_{ab}$ arises from a diffeomorphism covariant action and therefore satisfies a contracted Bianchi identity $\nabla^b A_{ab}=0$. Linearizing around a background solution gives, for any $h_{ab}$,
\be
 \nabla^b A_{ab}^{(1)}[h]=0
\ee
so, when (\ref{harmoniceqofmotion}) is satisfied, the divergence of (\ref{Atdef}) gives
\be
\label{harmonicprop}
 \nabla^b \nabla_b H_a +R_{ab} H^b=0.
\ee
This is a standard linear wave equation so provided the initial data is chosen such that $H_a$ and its first time derivative vanish then the solution will have $H_a \equiv 0$. (As for the Einstein equation, vanishing of the first time derivative of $H_a$ is equivalent, via the equation of motion, to the condition that the initial data satisfies the constraint equations \cite{Choquet-Bruhat1988}.)  This proves that the gauge condition (\ref{harmonic}) is propagated by the equation of motion (\ref{harmoniceqofmotion}). Hence the resulting solution will satisfy the original equation of motion (\ref{linearized}). 

The harmonic gauge equation of motion (\ref{harmoniceqofmotion}) takes the form
\be
 P^{abcdef} \nabla_e \nabla_f  h_{cd} + \ldots = 0
\ee
where the ellipsis denotes terms involving fewer than two derivatives of $h_{ab}$. The coefficient here defines the principal symbol
\be
 P(\xi)^{abcd} \equiv P^{abcdef} \xi_e \xi_f
\ee
for an arbitrary covector $\xi_a$. The coefficient is symmetric in $ab$ and in $cd$. It can be split into the terms coming from the (harmonic gauge) Einstein tensor, and those coming from the extra Lovelock terms:
\be
P(\xi)^{abcd}  = P_{\rm Einstein}(\xi)^{abcd} + \delta P^{abcd}(\xi) 
\ee
where, for a symmetric tensor $t_{ab}$,
\be 
\label{einsteinsymbol}
 P_{\rm Einstein}(\xi)^{abcd} t_{cd} = -\frac{1}{2} \xi^2 G^{abcd} t_{cd} 
\ee
with $\xi^2 = g^{ab} \xi_a \xi_b$ and 
\be
\label{Gdef}
 G^{abcd} = \frac{1}{2} \left( g^{ac} g^{bd} + g^{ad} g^{bc} - g^{ab} g^{cd} \right).
\ee
Viewed as a quadratic form on symmetric tensors, $G^{abcd}$ has signature $(d,d(d-1)/2)$, i.e., $d$ negative eigenvalues and $d(d-1)/2$ positive eigenvalues. 

The Lovelock contribution to the principal symbol is given by \cite{Reall2014a}
\be
 \delta P^a{}_b{}^{cd}(\xi) t_{cd} \equiv \delta P^a{}_b{}^{cdef} \xi_e \xi_f t_{cd} = -2 \sum_{p \ge 2} p k_p \delta^{ace g_3 g_4 \ldots g_{2p-1} g_{2p}}_{bdf h_3 h_4 \ldots h_{2p-1} h_{2p}} t_{c}{}^{d} \xi_e \xi^f R_{g_3 g_4}{}^{h_3 h_4} \ldots R_{g_{2p-1} g_{2p}}{}^{h_{2p-1} h_{2p}}.
\ee
Note that 
\be
  \delta P^{abcdef}= \delta P^{cdabef}
\ee
and
\be
\label{gaugesyma}
 \delta P^{(a|bcd|ef)}=\delta P^{a(bc|de|f)}=0.
\ee
These identities are a consequence of the the gauge symmetry of the theory and the fact that the gauge fixing terms do not affect $\delta P$. We will discuss this in more detail in section \ref{sec:gaugenull}. It follows that
\be
\label{gaugesym}
 \xi_a \delta P^{abcd} (\xi)= \xi_b \xi_c \xi_f \delta P^{abcdef} = 0.
\ee

\subsection{Setting up the problem}

\label{sec:setup}

We will investigate whether the harmonic gauge linearized Lovelock equation of motion is hyperbolic when the curvature of the background spacetime is small. Here, ``small'' means small compared to any of the scales defined by the dimensionful coupling constants $k_p$, so one expects the Lovelock terms in the equation of motion to be small compared to the Einstein term. 

To relate to the discussion of (\ref{sec:secondorder}) we need to introduce coordinates $x^\mu = (t,x^i)$. We assume that these are chosen so that surfaces of constant $t$ are spacelike, i.e. $g^{00}<0$, which ensures that the initial value problem for the harmonic gauge linearized Einstein equation is well-posed. We want to ask whether the same is true for the harmonic gauge linearized Lovelock equation when the background curvature is small. Here, by small, we mean that there exists an orthonormal basis $\{e_\mu^a\}$ with $e_0$ orthogonal to surfaces of constant $t$, for which the magnitude of the largest component of the Riemann tensor is $L^{-2}$ where $|k_p| L^{-2p} \ll 1$ for all $p\ge 2$. This ensures that the Lovelock terms in the equation of motion are small compared to the Einstein term. 

The principal symbol $P(\xi)$ maps symmetric tensors to symmetric tensors so we regard it as a $N \times N$ matrix where $N = d(d+1)/2$. We define $N \times N$ matrices $A(x)$, $B(x,\xi_i)$ and $C(x,\xi_i)$ using equation (\ref{ABCdef}), i.e.,
\be
 A = P^{00} \qquad B(\xi_i) = 2 \xi_i P^{0i} \qquad C(\xi_i) = \xi_i \xi_j P^{ij}.
\ee 
Here $\xi_i$ is real with $\xi_i \xi_i =1$ (since this is what we need in the definitions of strong and weak hyperbolicity). Throughout this section we will not write explicitly the dependence on the spacetime coordinates $x^\mu$. Note that these matrices are real and symmetric: the latter property arises because the equation of motion can be obtained from a Lagrangian (see section \ref{sec:gaugenull}). 

Our assumption that the surfaces of constant $t$ are spacelike ensures that $A$ is invertible when the Lovelock terms are absent. Hence, by continuity, $A$ is also invertible when the background curvature is small. We can therefore define $M(\xi_i)$ as above, i.e., 
\be
\label{MdefLovelock}
 M(\xi_i) = \begin{pmatrix} 0 & I \\ -A^{-1} C(\xi_i) & - A^{-1}B(\xi_i) \end{pmatrix}.
\ee
Recall that weak hyperbolicity is the requirement that the eigenvalues of this matrix are real. For strong hyperbolicity it is necessary that the eigenvalues are real and the matrix is diagonalizable.

From the discussion of section (\ref{sec:secondorder}) we know that $\xi_0$ is an eigenvalue of $M(\xi_i)$ if, and only if, the corresponding eigenvector $v$ has the form 
\be
\label{vdef}
 v= \begin{pmatrix} t\\  \xi_0 t \end{pmatrix} 
\ee
for some non-zero symmetric $t_{\mu\nu}$ such that
\be
P(\xi) t = 0
\ee 
where $\xi_\mu \equiv (\xi_0,\xi_i)$ in the argument of $P$. 

Consider first the case of the linearized Einstein equation. Since $G^{abcd}$ is non-degenerate, equation (\ref{einsteinsymbol}) implies that $\xi_\mu$ is characteristic if, and only if, it is null:
\be
 P_{\rm Einstein}(\xi) t =0, \qquad t\ne 0 \qquad \Leftrightarrow \qquad g^{\mu\nu} \xi_\mu\xi_\nu=0.
\ee
Let $\xi_0^\pm$ denote the two solutions of $g^{\mu\nu} \xi_\mu \xi_\nu=0$ for the given $\xi_i$. Of course these solutions are real, so the (harmonic gauge) Einstein equation is weakly hyperbolic. We define the null covectors
\be
\xi_\mu^\pm=(\xi_0^\pm,\xi_i).
\ee
These covectors will play an important role throughout this paper. By solving explicitly one finds that
\be
 \xi_0^+ + \xi_0^- = -2 \frac{g^{0i} \xi_i}{g^{00}} \qquad \Rightarrow \qquad \xi^{0+} + \xi^{0-} = 0.
\ee 
Hence we can adopt the convention $\xi^{0+}<0$, $\xi^{0-}>0$.\footnote{\label{xi0}
We cannot have $\xi^{0\pm}=0$ because that would violate the facts that $\xi_\mu^\pm$ is null and $e^0$ is timelike.} 

We have $P_{\rm Einstein}(\xi^\pm)t=0$ for {\it any} $t_{ab}$. Hence for the Einstein equation, the matrix $M$ has two real eigenvalues $\xi_0^\pm$ and the associated eigenvectors are $(t,\xi_0^\pm t)^T$. Each eigenvalue has $N$ eigenvectors associated to it. It follows that $M$ has $2N$ linearly independent eigenvectors and hence $M$ is diagonalizable, as required by strong hyperbolicity. 

We now return to the general case of Lovelock theory. Define a $2N \times 2N$ real symmetric (and hence Hermitian) matrix $H(\xi_i)$ by
\be
\label{HdefLovelock}
 H(\xi_i) =  \begin{pmatrix} B(\xi_i) & A \\ A &0 \end{pmatrix} .
\ee
We then have
\be
\label{HMeq}
 H(\xi_i) M(\xi_i) H(\xi_i)^{-1} = M(\xi_i)^T
\ee
so $M$ is real symmetric (and hence Hermitian) w.r.t. $H$. It is easy to see that $H$ is non-degenerate: if $v=(t,t')^T$ then $Hv=0$ implies $t=t'=0$ using the fact that $A$ is invertible.\footnote{The matrix $H$ is closely related to the symplectic current density $\omega^\mu$ defined in \cite{Lee1990}. Roughly speaking, $H$ is the high spatial frequency part of the Fourier space analogue of $-i\omega^0$.} $H$ is Hermitian and non-degenerate so its eigenvalues are real and non-zero. We can determine the signature of $H$ by writing the Lovelock couplings as 
\be
\label{epsilondef}
k_p = \epsilon \tilde{k}_p \qquad p \ge 2 .
\ee
Since the eigenvalues of $H$ are real, non-vanishing, and depend continuously on $\epsilon$ (with $\tilde{k}_p$ and the background curvature fixed), the signature of $H$ cannot depend on $\epsilon$. Hence it can be evaluated at $\epsilon=0$, i.e. for the linearized Einstein equation. The result is that $H$ has $N$ positive eigenvalues and $N$ negative eigenvalues, even for strong background fields. Thus, although $H$ and $M$ satisfy the condition (\ref{Hcond2}), this does not imply strong hyperbolicity because $H$ is not positive definite.\footnote{This is the case even for the Einstein equation ($\epsilon=0$). However, for the Einstein equation we have shown that we can diagonalize $M$ so we can construct a positive definite matrix $K$ as explained above equation (\ref{Hcond}).}

\subsection{Proof of weak hyperbolicity in a low curvature background}

\label{sec:weakproof}

To proceed, we will use a continuity argument involving the parameter $\epsilon$ defined in (\ref{epsilondef}). Note that taking $\epsilon$ small at fixed $\tilde{k}_p$ and fixed background curvature is equivalent to assuming the background curvature to be small at fixed $k_p$. We will establish weak hyperbolicity for small $\epsilon$, which is equivalent to establishing it for small background curvature. In what follows we will suppress the dependence of $M$ and $H$ on $\xi_i$ and write simply $M(\epsilon)$ and $H(\epsilon)$. 

For $\epsilon=0$ we showed above that $\xi_0^\pm$ are the only eigenvalues of $M(\epsilon)$, each with degeneracy $N$. The eigenvalues of $M(\epsilon)$ depend continuously on $\epsilon$ \cite{Kato1976}. Hence, for small $\epsilon$, they can be split unambiguously into two sets according to whether they approach $\xi_0^+$ or $\xi_0^-$ as $\epsilon \rightarrow 0$. We will follow \cite{Kato1976} and refer to these sets as the {\it $\xi_0^+$-group} and the {\it $\xi_0^-$-group}. Each group contains $N$ eigenvalues. 

Since we do not know whether or not the eigenvalues and eigenvectors of $M(\epsilon)$ are real, we will regard $M(\epsilon)$ and $H(\epsilon)$ as acting on a {\it complex} vector space $V$ of dimension $2N$. 

For $\epsilon=0$, the eigenvalues $\xi_0^\pm$ are degenerate but ``semi-simple'', i.e., $M(0)$ is diagonalizable. However, there is no reason for this to remain true when $\epsilon \ne 0$: the Jordan canonical form of $M(\epsilon)$ may involve non-trivial Jordan blocks. For any eigenvalue $\xi_0$, one can define a {\it generalized eigenspace} as
\be
\label{generalizedespace}
 \left\{ v: \exists r\;  {\rm such \; that} \; (M-\xi_0 I)^r v = 0\right \}.
\ee
This is the sum of the vector spaces associated with the Jordan blocks corresponding to that eigenvalue. We define the ``total generalized eigenspace for the $\xi_0^\pm$-group'' $V^\pm(\epsilon)$ as the sum over generalized eigenspaces of the eigenvalues in the $\xi_0^\pm$-group. Since any eigenvalue belongs to one of these groups we have
\be
 V = V^+(\epsilon) \oplus V^-(\epsilon) .
\ee
We denote the projection onto $V^\pm(\epsilon)$ as $\Pi^\pm(\epsilon)$, i.e., 
\be
 V^\pm(\epsilon) = \Pi^\pm(\epsilon) V .
\ee
These projection matrices are holomorphic in $\epsilon$ for small $\epsilon$, in fact there is an explicit formula \cite{Kato1976}
\be
 \Pi^\pm (\epsilon) = -\frac{1}{2\pi i} \int_{\Gamma^\pm} (M(\epsilon)-z)^{-1} dz 
\ee
where $\Gamma^\pm$ is a simple closed curve in the complex plane such that $\xi_0^\pm$ lies inside $\Gamma^\pm$ but $\xi_0^\mp$ lies outside $\Gamma^\pm$. Note that $\Gamma^\pm$ does not depend on $\epsilon$. For small non-zero $\epsilon$, the integrand has poles at the eigenvalues of $M(\epsilon)$ but only the eigenvalues that belong to the $\xi_0^\pm$-group lie inside $\Gamma^\pm$.

It can be shown that $M(\epsilon)$ and $H(\epsilon)$ satisfying \eqref{HMeq} can be brought simultaneously to a block-diagonal canonical form, where $M(\epsilon)$ is in Jordan canonical form and $M(\epsilon)$ and $H(\epsilon)$ have the same block structure \cite{Gohberg1983}. Since $V^+(\epsilon)$ and $V^-(\epsilon)$ contain different Jordan blocks of $M(\epsilon)$ it follows that these subspaces are orthogonal w.r.t. $H(\epsilon)$. Consider the restriction of $H(\epsilon)$ to these subspaces. Define the projection of $H(\epsilon)$ onto $V^\pm(\epsilon)$:
\be
 H^\pm(\epsilon) = \Pi^\pm(\epsilon)^\dagger H(\epsilon) \Pi^\pm(\epsilon).
\ee
This is a Hermitian matrix which depends holomorphically on $\epsilon$. We will need to determine its signature. 
Any vector in $V^\mp(\epsilon)$ is an eigenvector with eigenvalue $0$ hence $H^\pm(\epsilon)$ has at least $N$ vanishing eigenvalues. The remaining eigenvalues are associated to eigenvectors living in $V^\pm(\epsilon)$. Since the restriction of $H^\pm(\epsilon)$ to $V^\pm$ is the same as the restriction of $H(\epsilon)$ to $V^\pm$, it follows that this restriction is non-degenerate, i.e., these remaining eigenvalues are all non-zero. Therefore we can determine the signs of these eigenvalues by looking at the signs of the eigenvalues when $\epsilon=0$, and using continuity. For $\epsilon=0$, we know that $V^\pm(0)$ consists of vectors of the form $v=(t,\xi_0^\pm t)^T$. Taking the inner product of two such vectors w.r.t. $H^\pm(0)$ gives
\be
\label{v1v2}
 v_1^\dagger H^\pm(0) v_2 = t_1^\dagger B(0) t_2 + 2 \xi_0^\pm t_1^\dagger  A(0) t_2 =2 \xi_\mu^\pm t_1^\dagger P^{0\mu}_{\rm Einstein} t_2  = -\xi^{0\pm} t_1^\dagger G t_2
\ee
where $G$ is defined in (\ref{Gdef}). 
Hence the signature of $H^\pm(0)$ restricted to $V^\pm(0)$ is the same as the signature of $-\xi^{0\pm} G$. Recall that $\xi^{0+}<0$, $\xi^{0-}>0$. It follows that within $V^\pm(0)$, $H^\pm(0)$ has the same signature as $\pm G$, i.e., $d$ negative eigenvalues and $d(d-1)/2$ positive eigenvalues for $H^+(0)$ and vice-versa for $H^-(0)$. Hence, by continuity, it follows that  $H^+(\epsilon)$ has $d$ negative eigenvalues and $d(d-1)/2$ positive eigenvalues, with eigenvectors in $V^+(\epsilon)$, as well as $N=d(d+1)/2$ vanishing eigenvalues with eigenvectors in $V^-(\epsilon)$. Similarly for $H^-(\epsilon)$ with positive and negative interchanged. 

We can identify an important subset of eigenvectors of $M(\epsilon)$ explicitly, for any $\epsilon$. They are associated to a residual gauge freedom. These ``pure gauge'' eigenvectors have $v$ of the form (\ref{vdef}) with
\be
\xi_0 = \xi_0^\pm \qquad t_{\mu\nu} = \xi^\pm_{(\mu} X^{}_{\nu)}
\ee
for arbitrary complex $X_\mu$. Of course a pure gauge eigenvector with eigenvalue $\xi_0^\pm$ belongs to $V^\pm(\epsilon)$. It is interesting to calculate the inner product of two pure gauge eigenvectors so let $t'_{\mu\nu} = \xi^\pm_{(\mu} X'_{\nu)}$ and consider the associated vector $v'$ defined by (\ref{vdef}). Since $v,v'$ are elements of $V^\pm(\epsilon)$, their inner product w.r.t. $H^\pm(\epsilon)$ is the same as their inner product w.r.t. $H(\epsilon)$:
\be
\label{gaugenull}
 v^{'\dagger} H(\epsilon) v = t^{'\dagger} B(\epsilon) t + 2 \xi_0^\pm t^{'\dagger} A(\epsilon) t = 2 \xi_\mu^\pm t^{'\dagger} P^{0\mu}(\epsilon) t= 2 \xi_\mu^\pm \xi_\nu^\pm \xi_\rho^\pm \bar{X}'_\sigma X_\tau P^{\nu\sigma \rho \tau 0 \mu}(\epsilon) =0 
\ee
where in the final step we used the second equation in (\ref{gaugesym}), and the fact that two such ``pure gauge'' vectors $t,t'$ are orthogonal w.r.t. $G^{\mu\nu\rho\sigma}$. This result shows that the pure gauge eigenvectors with eigenvalue $\xi_0^\pm$ form a $d$-dimensional subspace ${N}^\pm$ of $V^\pm(\epsilon)$ that is null w.r.t. $H^\pm(\epsilon)$. 

We can now prove that the harmonic gauge linearized equation of motion of Lovelock theory is weakly hyperbolic in a small curvature background. 
Consider the possibility of an eigenvalue $\xi_0$ that is complex, with eigenvector $v$. For concreteness, assume that $\xi_0$ belongs to the $\xi_0^+$-group, so $v \in V^+(\epsilon)$. Equation \eqref{HMeq} implies that a pair of eigenvectors whose eigenvalues are not complex conjugates of each other must be orthogonal w.r.t. $H(\epsilon)$. This implies that $v$ is orthogonal, w.r.t. $H^+(\epsilon)$, to the ``pure gauge'' eigenvectors in $V^+(\epsilon)$. Furthermore, since $\xi_0$ is complex, the $H(\epsilon)$-norm of $v$ must vanish, which implies that $v$ is null w.r.t. $H^+(\epsilon)$. The linear span of $v$ and $N^+$ now gives a $(d+1)$-dimensional subspace of $V^+(\epsilon)$ that is null w.r.t. $H^+(\epsilon)$. However, this is impossible because we showed above that for small $\epsilon$, $H^+(\epsilon)$ has $d$ negative eigenvalues and $d(d-1)/2$ positive eigenvalues which implies the maximal dimension of a null subspace of $V^+(\epsilon)$ is given by ${\rm min}(d,d(d-1)/2)=d$ \cite{Gohberg1983}. This proves that complex $\xi_0$ is not possible for small $\epsilon$.

The final step is to note that the above argument assumed fixed $\xi_i$, i.e., for given $\xi_i$ then complex $\xi_0$ is not possible for small enough $\epsilon$. But we need our final result to be {\it uniform} in $\xi_i$, i.e., we need to show that the upper bound on $\epsilon$ does not depend on $\xi_i$. To do this we recall that our definition of weak hyperbolicity refers only to $\xi_i$ satisfying the condition $\xi_i \xi_i = 1$, i.e., $\xi_i$ belonging to a compact set. The spectrum of a matrix $M$ has uniformly continuous dependence on $M$ when $M$ is restricted to a bounded set \cite{Kato1976}. It follows that the spectrum of $M(\epsilon)$ and $H(\epsilon)$ has uniformly continuous dependence on $\epsilon$ and $\xi_i$ when $\epsilon$ is restricted to a bounded set and $\xi_i \xi_i=1$. Using this it can be shown that our results above are indeed uniform in $\xi_i$. The same argument establishes that our result is uniform in the spacetime point $x^\mu$ provided we restrict to a compact region of spacetime.  

The above argument is restricted to a weakly curved background spacetime. If the curvature is not weak then the argument can fail. Imagine increasing $\epsilon$ to arbitrarily large values. There are two things that could go wrong. First, our assumption that $A$ is invertible may fail, i.e., we might reach a value of $\epsilon$ for which a surface of constant $t$ becomes characteristic somewhere. Second, it might not be possible to separate the eigenvalues into the $\xi_0^+$ group and the $\xi_0^-$ group as we did above. For example, as we increase $\epsilon$, an eigenvalue from one group might coincide with an eigenvalue from the other group. At larger $\epsilon$, this eigenvalue could then split into a pair of complex conjugate eigenvalues, violating weak hyperbolicity.

\subsection{Failure of strong hyperbolicity in a generic low curvature background}

\label{sec:Lovelock_strong}

For strong hyperbolicity, $M$ must be diagonalizable. We will now demonstrate that this is not the case for a {\it generic} weakly curved background spacetime.\footnote{In this section, we will not write explicitly the dependence on the parameter $\epsilon$ e.g. we write $M$ instead of $M(\epsilon)$.} We showed above that eigenvalues $\xi_0$ are all real in a weakly curved background. Therefore in this section we will assume that all vector spaces $V^\pm$, $N^\pm$, etc., are real. Note that the assumption that the background is weakly curved is required to define these spaces. 

As discussed above, $M$ and $H$ satisfying (\ref{HMeq}) can be brought simultaneously via a change of basis to a certain canonical form \cite{Gohberg1983}. We need to discuss this canonical form in more detail. In the canonical basis, $M$ has Jordan normal form and $H$ is block diagonal, with the same block structure as $M$. By this we mean that a $n \times n$ Jordan block in $M$ corresponds to a $n \times n$ block in $H$. Such a block of $H$ consists of zeros everywhere except on the diagonal running from top right to bottom left. Along this diagonal, the elements are all equal to $1$ or all equal to $-1$. For example, if $M$ has a $3 \times 3$ Jordan block then the corresponding $3 \times 3$ block in $H$ has the form
\be
\label{Hblock3}
  \begin{pmatrix}  0 & 0 & \pm 1 \\ 0 & \pm 1 & 0 \\ \pm 1 & 0 & 0 \end{pmatrix}.
\ee
Each $n \times n$ block in $H$ is non-degenerate and has signature either $+1$ or $-1$ (if $n$ is odd) or $0$ (if $n$ is even). 

Recall the definition  (\ref{generalizedespace}) of a generalized eigenspace. Note that a generalized eigenspace corresponds to a sum of all Jordan blocks associated to the given eigenvalue. Hence $V^\pm$ is a direct sum of the basis vectors associated to Jordan blocks of eigenvalues in the $\xi_0^\pm$-group. Hence any Jordan block is associated either to $V^+$ or to $V^-$. The canonical form then implies that $V^+$ and $V^-$ are orthogonal w.r.t. $H$, as stated above. 

Let $E^\pm \subset V^\pm$ denote the generalized eigenspace of the eigenvalue $\xi_0^\pm$. We have shown that $N^\pm \subset E^\pm$. Hence, when restricted to $E^\pm$, $H^\pm$ must admit a $d$-dimensional null subspace. Consider $H^+$. From the canonical form we know that $H^+$ is non-degenerate when restricted to $E^+$. If this restriction has signature $(r,s)$, then the dimension of a maximal null subspace of $E^+$ is $\min(r,s)$  \cite{Gohberg1983}; hence, we have $r,s \ge d$. However we already know that $H^+$ has signature $(d,d(d-1)/2)$ within $V^+$. The canonical form for $H$ shows that the signature is equal to the union of the signatures of each block. Therefore $H^+$ can have at most $d$ negative eigenvalues within $E^+$, i.e., we must have $r \le d$. Combining these inequalities we see that $r=d$ and $s \ge d$. Hence $E^+$ has dimension $r+s \ge 2d$. Similarly $E^-$ has dimension at least $2d$. 

A necessary condition for strong hyperbolicity is that $M$ is diagonalizable, i.e. there should be no non-trivial Jordan blocks. In other words, strong hyperbolicity requires that all generalized eigenspaces are simply eigenspaces. Hence if the theory is strongly hyperbolic then $E^\pm$ must be an eigenspace. Hence strong hyperbolicity requires that $M$ admits at least $2d$ eigenvectors with eigenvalue $\xi_0^\pm$. We already know that there are $d$ such eigenvectors in $N^\pm$. But for strong hyperbolicity there must exist at least an extra $d$ eigenvectors beyond these ``pure gauge'' ones. In terms of the principal symbol, this condition is that there exist at least $2d$ solutions $t_{ab}$ of $P(\xi^\pm) t = 0$ or equivalently (since $P_{\rm Einstein}(\xi^\pm)=0$) $\delta P(\xi^\pm) t = 0$. In other words $\ker \delta P(\xi^\pm)$ should have dimension at least $2d$. Furthermore, for strong hyperbolicity, this must be true for any $\xi_i$ and hence for any null $\xi^\pm$. In other words:

\medskip

\noindent {\it A necessary condition for strong hyperbolicity is that, for any null $\xi$, $\ker \delta P(\xi)$ has dimension at least $2d$.} 

\medskip

There are certainly examples of background spacetimes for which this condition is satisfied. An extreme example is a flat background, for which $\delta P=0$. In this case $M$ is diagonalizable and the equation of motion is strongly hyperbolic. A less trivial example is supplied by the class of Ricci flat spacetimes with Weyl tensor of type N, which are solutions of Lovelock theory with $\Lambda=0$. In this case, the results of Ref.~\cite{Reall2014} imply that $M$ is diagonalizable so the equation of motion is strongly hyperbolic in such a background (even for large curvature). For this class of spacetimes, in addition to the pure gauge eigenvectors, generically there exist $d$ additional eigenvectors in $E^\pm$. This implies that $\ker \delta P(\xi^\pm)$ generically has dimension $2d$ for these spacetimes, in agreement with the above argument.

These background spacetimes are clearly very special because they have symmetries. In a {\it generic} weakly curved background, with null $\xi$, there is no reason to expect that $\ker \delta P(\xi)$ contains any non-gauge elements. To explain this, first note that if we are interested in non-gauge elements of $\ker \delta P(\xi^\pm)$ then we can regard $\delta P(\xi^\pm)$ as a map from the quotient space $V^\pm / N^\pm$, which has dimension $d(d-1)/2$, to the space of symmetric tensors which have vanishing contraction with $\xi^\pm$ (because of (\ref{gaugesym})). The latter space also has dimension $d(d-1)/2$. There is no reason to expect this map to have non-trivial kernel. 

Perhaps we are overlooking some hidden symmetry of $\delta P$ that would guarantee that its kernel is larger than we expect. To exclude this possibility, we have calculated $\ker \delta P(\xi)$ for null $\xi$ in a generic background using computer algebra as follows. We fix a point in spacetime and work at that point. Note that $\delta P$ is determined by the Riemann tensor of the background. For given null $\xi$ we can introduce a null basis for which $\xi$ is one of the basis vectors. In this basis, we can generate a random Riemann tensor satisfying the background equation of motion. To do this, we generate a random (small) Weyl tensor and then use the background equation of motion to determine the Ricci tensor and hence the Riemann tensor. Since the equation of motion is nonlinear in curvature, there can be multiple solutions for the Ricci tensor but typically only one of these has small components, so this is the one we use. We then calculate $\ker \delta P(\xi)$ for this background Riemann tensor. The result is that, generically, this kernel has dimension $d$, i.e., it consists only of the ``pure gauge'' elements. 
\medskip

In summary, we have shown: $M$ is not diagonalizable for a generic weak field background. Therefore {\it the harmonic gauge linearized Lovelock equation of motion is not strongly hyperbolic in a generic weak field background.} 
\medskip

It is interesting to consider the canonical form of $M$ in more detail. Let's examine the condition for $M$ to have a $n \times n$ Jordan block with $n \ge 2$. From the canonical form, it is clear that the eigenvector associated to such a block must be null.\footnote{For example, for a $3 \times 3$ block, in the canonical basis, the eigenvector is $(1,0,0)^T$ and evaluating the norm of this using (\ref{Hblock3}) gives $0$.} Assume that this eigenvector lives in $V^+$. If the eigenvalue is not $\xi_0^+$ then this eigenvector must be $H^+$-orthogonal to $N^+$, which implies that we could add this eigenvector to $N^+$ to construct a null subspace of dimension $d+1$, contradicting the fact that $N^+$ is a maximal null subspace. Hence the eigenvalue must be $\xi_0^+$. Similarly if the eigenvector lives in $V^-$ then the eigenvalue is $\xi_0^-$. We conclude that a non-trivial Jordan block must have eigenvalue $\xi_0^\pm$, so the basis vectors associated to the block must lie in $E^\pm$. 

Any such Jordan block admits a vector $v \in E^\pm$ such that $(M-\xi_0^\pm)^2 v = 0$ but $(M-\xi_0^\pm) v \ne 0$ ($v$ is simply the second basis vector associated to the block) hence $(M-\xi_0^\pm)v$ is an eigenvector of $M$ with eigenvalue $\xi_0^\pm$. So we must have
\be
\label{2ndorder}
 \left( M - \xi_0^\pm \right) v = \begin{pmatrix} s \\ \xi^\pm_0 s \end{pmatrix}
\ee
for some non-zero $s_{\mu\nu}$ such that (using $P_{\rm Einstein}(\xi^\pm)=0$)
\be
\label{Ps}
 \delta P(\xi^\pm) s = 0 \,.
\ee
To examine whether such a block is possible, we need to determine whether (\ref{2ndorder}) admits a solution $v$ for some $s_{\mu\nu} \ne 0$. If such a solution exists then $M$ is not diagonalizable. 

Writing $v=(t,t')^T$ we find that (\ref{2ndorder}) reduces to 
 \be
 \label{2ndorder1}
  t'=\xi_0^\pm t+ s 
\ee
and 
\be
\label{2ndorder2}
 \delta P(\xi^\pm) t = -(2\xi_0^\pm A + B) s .
\ee
The necessary and sufficient condition for this equation to admit a solution $t$ is for the RHS to have vanishing contraction with any element of $\ker \delta P(\xi^\pm)$. We know this kernel always contains the ``pure gauge'' eigenvectors, i.e., it contains $N^\pm$. So contract with a ``pure gauge'' vector of the form $r_{\mu\nu} = \xi^\pm_{(\mu} Y^{}_{\nu)}$. The LHS vanishes and we can rewrite the RHS in terms of $H$ to obtain
\be
 0 = (r, \xi^\pm_0 r) H   \begin{pmatrix} s \\ \xi^\pm_0 s \end{pmatrix} .
\ee
Hence $(s,\xi_0^\pm s)^T$ must be orthogonal (w.r.t. $H$) to all pure gauge eigenvectors in $E^\pm$, i.e., orthogonal to $N^\pm$. Furthermore, equation (\ref{Ps}) shows that $s$ belongs to the kernel of $\delta P(\xi^\pm)$ so we also need the contraction of $s$ with the RHS of (\ref{2ndorder2}) to vanish. This implies that $(s,\xi_0^\pm s)^T$ is null w.r.t. $H$. Therefore if this vector is not pure gauge, we could add it to $N^\pm$ to enlarge our null subspace, contradicting maximality of this null subspace. This proves that $s$ must be pure gauge, i.e., 
\be
 s_{\mu\nu} = \xi^\pm_{(\mu} X^{}_{\nu)}
 \ee
for some $X_\nu \ne 0$. Hence, {\it non-trivial Jordan blocks can arise only from pure gauge eigenvectors}. For $s_{\mu\nu}$ of this form, the RHS of (\ref{2ndorder2}) has vanishing contraction with any element of $N^\pm$. 

We argued above that, in a generic weakly curved background, {\it all} elements of $\ker \delta P(\xi^\pm)$ are ``pure gauge'', i.e., $\ker \delta P(\xi^\pm)=N^\pm$. It follows that in such a background, (\ref{2ndorder2}) can be solved for any pure gauge $s_{\mu\nu}$, i.e., {\it all} pure gauge eigenvectors belong to non-trivial Jordan blocks of $M$. So generically there are $d$ non-trivial Jordan blocks in each of $E^\pm$ and $M$ has $2d$ non-trivial blocks in total. In non-generic backgrounds, $\ker \delta P(\xi^\pm)$ may contain non-gauge elements in which case $M$ may have fewer than $2d$ non-trivial blocks. 

We have shown that, in a generic weak field background, every pure gauge eigenvector is associated to a $n \times n$ Jordan block of $M$ with $n \ge 2$. It is interesting to ask whether we could have $n \ge 3$. If $n \ge 3$ then there is a vector $v \in E^\pm$ such that $(M-\xi_0^\pm)^3 v = 0$ with $(M-\xi_0^\pm)^2 v \ne 0$. Let $(M-\xi_0^\pm)v \equiv (t,t')^T$, then $(t,t')$ must obey the equations (\ref{2ndorder1}), (\ref{2ndorder2}). Writing $v=(u,u')^T$ then gives
\be
 u' = \xi_0^\pm u + t
\ee
\be
 \delta P(\xi^\pm) u =  -(2\xi_0^\pm A + B) t - A s .
\ee
As with (\ref{2ndorder2}), the necessary and sufficient condition for this equation to admit a solution is that the RHS has vanishing contraction with any element of $\ker  \delta P(\xi^\pm)$. Generically we have $\ker  \delta P(\xi^\pm)=N^\pm$ so we need the RHS to have vanishing contraction with any pure gauge vector $r_{\mu\nu} = \xi^\pm_{(\mu} Y^{}_{\nu)}$. 
This contraction is just the $H$-inner product of $(t,t')$ with $(r,\xi_0^\pm r)$, so these vectors must be $H$-orthogonal for any pure gauge vector $r$. But there is no reason why this should be true. So generically we do not expect the above equations to admit a solution, i.e., the generic situation is $n=2$. 

To summarize: we have shown that, in a generic weak field background, every pure gauge eigenvector of $M$ belongs to a Jordan block of size $2\times 2$.\footnote{
More precisely, this is true for a generic point and for generic $\xi_i$, in a generic weakly curved background.} Since non-trivial Jordan blocks can arise only from pure gauge eigenvectors, it follows that, generically, $V^\pm$ consists of $d$ $2\times 2$ Jordan blocks, one for each pure gauge eigenvector, and $d(d-3)/2$ additional non-gauge eigenvectors. For a generic Ricci flat type N spacetime, it has been shown that these $d(d-3)/2$ additional eigenvectors have eigenvalues distinct from $\xi_0^\pm$ \cite{Reall2014} so they do not belong to $E^\pm$ hence we expect this to be the behaviour in a generic spacetime. Therefore, generically, $E^\pm$ will have dimension $2d$. 

Note that the $d(d-3)/2$ eigenvectors in $V^\pm$ that do not belong to $E^\pm$ can be regarded as the ``physical graviton polarizations''  \cite{Reall2014}. To understand why, note that these eigenvectors have the form (\ref{vdef}) where $t_{\mu\nu}$ satisfies the harmonic gauge condition. To prove the latter statement, simply contract the equation 
\be
 P(\xi)^{\mu\nu\rho\sigma} t_{\rho\sigma} = 0
\ee
with $\xi_\nu$ and use (\ref{gaugesym}) to obtain
\be
 \xi^2 \left( \xi^\nu t_{\mu\nu} - \frac{1}{2} \xi_\mu t^\rho_\rho \right)= 0 \qquad \Rightarrow \qquad  \xi^\nu t_{\mu\nu} - \frac{1}{2} \xi_\mu t^\rho_\rho =0
\ee
where we used the fact that $\xi^2\ne 0$ because the eigenvector is not in $E^\pm$. Here the LHS is the ``high frequency part'' of the harmonic gauge condition. It is easy to check that the ``pure gauge'' eigenvectors in $N^\pm$ also satisfy this condition. However, there is no reason to expect that the vectors $t_{\mu\nu}$ obtained by solving (\ref{2ndorder2}) will satisfy this condition. Hence, generically, the $d$ ``non-gauge'' vectors in $E^\pm$ are associated to $t_{\mu\nu}$ which violate the harmonic gauge condition. So generically $E^\pm$ consists only of ``pure gauge'' and ``gauge violating'' vectors, which is why the $d(d-3)/2$ elements of $V^\pm$ that do not belong to $E^\pm$ can be regarded as the ``physical polarizations''. 

\subsection{Dynamical violation of weak hyperbolicity} 

We have shown that the linearized harmonic gauge equation of motion of Lovelock theory is not strongly hyperbolic in a {\it generic} weak curvature background. However, as mentioned above, it can be strongly hyperbolic in a non-generic weak curvature background. In this section, we will discuss a class of such backgrounds, namely homogeneous, isotropic, cosmological solutions of Lovelock theory. The aim is to demonstrate that weak (and hence also strong) hyperbolicity can be violated {\it dynamically}: there are ``collapsing universe'' solutions that start with small curvature but develop large curvature over time, in such a way that weak hyperbolicity is violated. Once this happens, local well-posedness of the equation of motion is lost, which implies that generic linear perturbations of the solution can no longer be evolved. 

Lovelock theories admit FLRW-type solutions \cite{Wheeler1986,Deruelle1990}
\begin{equation}
	g=-dt^{2}+a(t)^{2}\gamma
\end{equation}
where \(\gamma\) is the metric of a \((d-1)\)-dimensional submanifold of constant curvature \(K\). We denote by \(D\) the Levi-Civita connection associated to \(\gamma\). The non-vanishing components of the Riemann tensor associated to \(g\) are
\begin{equation}
	R_{ij}{}^{kl}=\alpha(t)\delta^{kl}_{ij} \qquad R_{0i}{}^{0j}=\beta(t)\delta^{j}_{i}
\end{equation}
where, in terms of the Hubble parameter \(H=\dot{a}/a\),
\begin{equation}
	\alpha=\frac{K}{2a^{2}}+H^{2} \qquad \beta=H^{2}+\dot{H}.
\end{equation}
The non-vanishing components of the Lovelock tensor \eqref{eq:lovelock_tensor} are
\begin{align}
	A^{0}{}_{0}=&\sum_{p} k_{p}' \alpha^{p}\\
	A^{i}{}_{j}=&\delta^{i}_{j}\sum_{p} \frac{k_{p}'}{(d-1)}\alpha^{p-1}(2p \beta+(d-2p-1)\alpha)
\end{align}
where, for convenience, we have rescaled the coupling constants
\begin{equation}
	k_{p}'=2^{p}\frac{(d-1)!}{(d-2p-1)!}k_{p}\qquad k_{0}=\Lambda, \quad k_{1}=-1/4 \,.
\end{equation}
Taking our matter source to be a perfect fluid with equation of state \(P=\omega \rho\), the equations of motion read
\begin{align}
	\sum_{p} k_{p}' \alpha^{p} & =-\rho\\
	\beta & = - \frac{\sum_{p}k_{p}' \alpha^{p}\left[(d-1)(\omega+1)-2p\right]}{\sum_{p}2p k_{p}' \alpha^{p-1}}.
\end{align}
To observe how weak hyperbolicity can be violated dynamically in this setting, it is sufficient to look at the linearized equations for \emph{transverse-traceless tensor} perturbations \(g \rightarrow g+\delta g\):
\begin{equation}
	\delta g_{0\mu}=0 \quad \delta g_{ij}=2 a^{2} h_{ij} \quad h_{ij}=h_{ji} \quad \gamma^{ij}h_{ij}=0 \quad D^{i} h_{ij}=0.
\end{equation}
These are governed by the equation
\begin{equation}
	- F_{1}(t) \ddot{h}_{ij}+ F_{2}(t) a^{-2}(t) \, D_{k}D^{k} h_{ij} + \ldots=0
\end{equation}
where the ellipsis denotes terms with fewer than $2$ derivatives and we have defined
\begin{align}
 	F_{1}(t) & = \sum_{p} (d-3) p k_{p}' \alpha^{p-1}\\
 	F_{2}(t) & = \sum_{p}p k_{p}' [2(p-1)\alpha^{p-2}\beta + (d-2p-1)\alpha^{p-1}].
\end{align} 
From this we can read off the principal symbol (restricted to tensor perturbations) and construct the matrices \(A, B\) and \(C\) described in Section~\ref{sec:secondorder}
\begin{align}
	A^{ijkl} & = - \gamma^{i(k}\gamma^{l)j} F_{1}(t)\\
	B^{ijkl} & = 0\\
	C^{ijkl} & = \gamma^{i(k}\gamma^{l)j} a^{-2}(t) \gamma^{mn}\xi_{m}\xi_{n} F_{2}(t).
\end{align}
We can then compute the eigenvalues of \(M\), or equivalently find the \({\xi}_{0}\) that solves \(({\xi}_{0}^2 A+ C)t=0\). For \(F_{1}(t)\neq 0\) we find
\begin{equation}
	\xi_0 = \tilde{\xi}^{\pm}_{0}\equiv \pm \frac{1}{a(t)}\sqrt{\gamma^{ij}\xi_{i}\xi_{j}\frac{F_{2}(t)}{F_{1}(t)}} \,.
\end{equation}
Since \(\gamma\) is a Riemannian metric (hence it is positive definite), the hyperbolicity of the theory is determined by the sign of \(F_{2}(t)/F_{1}(t)\). If the background is weakly curved then the Einstein term dominates $F_1$ and $F_2$ and both of these quantities are negative
so $\tilde{\xi}^{\pm}_{0}$ are real and the theory is weakly hyperbolic. However, if the curvature becomes large, e.g. in a collapsing universe solution, then one of these quantities might become positive, which makes $F_2/F_1$ negative so the theory is no longer weakly hyperbolic. 

In agreement with the comments at the end of Section~\ref{sec:weakproof}, we see that weak hyperbolicity can fail either when $F_1$ vanishes, i.e., the matrix $A$ becomes singular, or when $F_2$ vanishes, in which case an eigenvalue from the $\xi_0^+$ group becomes equal to an eigenvalue from the $\xi_0^-$ group, i.e., it is no longer possible to distinguish these two groups.  

If $F_1$ or $F_2$ becomes positive then $\xi_0$ is imaginary and there exist linearized solutions which grow exponentially with time. For this reason, in the cosmology literature, a change in sign of $F_1$ or $F_2$ is usually referred to as an ``instability'' of the background solution. More specifically, if $F_1$ becomes negative then the background is said to suffer a ``ghost instability'' and if $F_2$ becomes negative it is said to suffer a ``gradient instability''.\footnote{This behaviour was first discussed in the context of cosmological solutions of Horndeski theories \cite{DeFelice2010,DeFelice2011,Kobayashi2011}.} However, this nomenclature is misleading. For the concept of stability to make sense, one needs the initial value problem for perturbations to be locally well-posed so that one can ask what happens when a generic initial perturbation is evolved in time. But when $F_1/F_2$ becomes negative then the equation for linear perturbations is not weakly hyperbolic which implies that the initial value problem is not well-posed: a generic linear perturbation cannot be evolved in time so dynamics no longer makes sense.\\
  
Further examples of dynamical violation of weak hyperbolicity can be obtained by considering the interior of a static, spherically symmetric black hole solution of a Lovelock theory \cite{Boulware1985,Wheeler1986}.  
For a large black hole, the equations for linear perturbations are weakly hyperbolic outside the event horizon \cite{Reall2014}.\footnote{We expect that they are also strongly hyperbolic although we have not checked this.} However, one can show that in the interior of  such a black hole, the equations of motion fail to be weakly hyperbolic in a region \(0<r<r_{*}\). Here \(r\) is the area-radius of the \((d-2)\)-spheres, orbits of the symmetry group. Inside the black hole, surfaces of constant $r$ are spacelike and $-\partial/\partial r$ provides a time orientation. One can impose initial data for linear perturbations on a surface $r=r_0>r_*$ inside the black hole. For large enough $r_0$, the curvature will be small on such a surface. Evolving this data then leads to a violation of weak hyperbolicity at time $r=r_*$. Generic linear perturbations cannot be evolved beyond this time. 

\section{Horndeski theories}

\label{sec:horndeski}

\subsection{Equations of motion}

Horndeski theories are the most general diffeomorphism covariant four-dimensional theories of gravity coupled to a scalar field, with second order equations of motion \cite{Horndeski1974}. The fields in such theories are the metric \(g\) and a scalar field \(\Phi\) and the equations of motion are obtained from an action of the form 
\begin{equation}
	S= \frac{1}{16\pi G}\int d^4 x \sqrt{-g} (\mathcal{L}_{1}+\mathcal{L}_{2}+\mathcal{L}_{3}+\mathcal{L}_{4}+\mathcal{L}_{5})
\end{equation}
where 
\begin{align}
	\LL_1&=R + X - V(\Phi)\\
	\LL_{2}&= \GG_{2}(\Phi,X)\\
	\LL_{3}&= \GG_{3}(\Phi,X) \square\Phi\\
	\LL_{4}&= \GG_{4}(\Phi,X)R+\partial_{X}\GG_{4}(\Phi,X)\delta^{ac}_{bd}\,\nabla_{a}\nabla^{b}\Phi \, \nabla_{c}\nabla^{d}\Phi\\
 	\LL_{5}&= \GG_{5}(\Phi,X)G_{ab}\nabla^{a}\nabla^{b}\Phi-\frac{1}{6}\partial_{X}\GG_{5}(\Phi,X)\delta^{ace}_{bdf}\,\nabla_{a}\nabla^{b}\Phi \, \nabla_{c}\nabla^{d}\Phi \, \nabla_{e}\nabla^{f}\Phi
\end{align}
and we have defined \(X=-\frac{1}{2}(\nabla \Phi)^{2}\). 

The term $\LL_1$ corresponds to Einstein gravity minimally coupled to a scalar field with potential $V(\Phi)$. We will refer to this theory as the Einstein-scalar-field theory. We assume that the functions $\GG$ depend smoothly on $\Phi$ and $X$. To eliminate degeneracies between the various terms (allowing for field redefinitions $\Phi \rightarrow \Phi'(\Phi)$) we will impose the following restrictions on these functions:
\be
\label{horn_cond}
 \GG_2(\Phi,0)=(\partial_X \GG_{2}) (\Phi,0) = \GG_3(\Phi,0)=\GG_4(0,0)=\GG_5(0,0)= 0.
\ee
The equations of motion for Horndeski theory are given by 
\begin{align}
	E^{ab}[g,\Phi] &\equiv  -\frac{1}{\sqrt{-g}} \frac{\delta S}{\delta g_{ab}} =0\\
	E_{\Phi}[g,\Phi] &\equiv -\frac{1}{\sqrt{-g}}\frac{\delta S}{\delta \Phi}=0.
\end{align}
To study the hyperbolicity of these equations, we linearize around a background solution \((g,\Phi)\), i.e. we consider \((g+h,\Phi+\psi)\) and linearize in \(h\) and \(\psi\)
\begin{align}
	E_{ab}[g+h,\Phi+\psi] & = E_{ab}[g,\Phi]+E^{(1)}_{ab}[h,\psi]+\ldots\\
	E_{\Phi}[g+h,\Phi+\psi] & = E_{\Phi}[g,\Phi]+E_{\Phi}^{(1)}[h,\psi]+\ldots
\end{align}
so the linearized equations of motion are
\be
\label{linhorndeski}
 E_{ab}^{(1)}[h,\psi]=E_{\Phi}^{(1)}[h,\psi]=0.
\ee
Recall that the equations of motion resulting from the Einstein-scalar-field theory are strongly hyperbolic if we impose the usual harmonic gauge condition which is\footnote{
More properly we should call this a Lorenz gauge condition, but we will refer to it as a harmonic gauge condition for the reasons discussed below equation \eqref{harmonic}.}
\be
 G^{abcd} \nabla_b h_{cd} \equiv \nabla_b h^{ab} - \frac{1}{2} \nabla^a h^b_b = 0
\ee
where $G^{abcd}$ is defined by \eqref{Gdef}. Motivated by this, we will attempt to obtain hyperbolic equations of motion for Horndeski theory by imposing a generalized harmonic gauge condition
\begin{equation}
	\label{eq:ghg}
	H_{a}\equiv (1+f) G_a{}^{bcd} \nabla_b h_{cd} -\mathcal{H}_{a}{}^{b}\nabla_{b}\psi=0
\end{equation}
where the scalar $f$ and the tensor ${\cal H}_a{}^b$ depend only on background quantities. The idea is that when we deform the theory away from the Einstein-scalar-field theory we may need to deform the gauge condition in order to preserve hyperbolicity. The quantities $f$ and ${\cal H}$ describe such a deformation.\footnote{Of course we could divide through by $(1+f)$ to absorb $f$ into ${\cal H}$. The reason for including $f$ here is that it leads to a more general class of gauge-fixed equations of motion when we perform the gauge-fixing procedure described below.} This gauge condition could be generalized further by including terms that do not involve derivatives of $h_{ab}$ or $\psi$. However such terms do not affect the principal symbol and therefore do not influence hyperbolicity. 

To see that we can impose such a gauge condition, let \(Y^{a}\) be a vector field and consider the infinitesimal diffeomorphism generated by $Y^a$: 
\be
\label{gauge_horn}
	h_{ab}  \rightarrow h_{ab}+\nabla_{(a}Y_{b)} \qquad \psi  \rightarrow \psi + Y\cdot \nabla\Phi.
\ee
Under such transformation \(H_{a}\) will change as
\begin{equation}
	H_{a}\rightarrow H_a+ \frac{1}{2}(1+f)(\nabla^{b}\nabla_{b}Y_{a}+R_{ab}Y^{b})-\mathcal{H}_{a}{}^{b}\nabla_{b}(Y\cdot\nabla\Phi)
\end{equation}
\(H_{a}\) can then be set to zero by choosing \(Y_{a}\) to solve
\begin{equation}
	\nabla^{b}\nabla_{b}Y_{a}- \frac{2}{1+f} \mathcal{H}_{a}{}^{b}\nabla_{b}(Y\cdot\nabla\Phi) + R_{ab} Y^{b}=
	-\frac{2}{1+f} H_a \,.
	\end{equation}
This is a linear wave equation of a standard type, which guarantees the existence of such \(Y_{a}\). Note that if we changed the way that the first derivatives of $h_{ab}$ appear in \eqref{eq:ghg} then this argument would no longer work. 

To obtain the equations of motion in the generalized harmonic gauge, consider expanding the action to quadratic order in $(h,\psi)$ to obtain an action governing the linearized perturbation. Now to this action we add the gauge-fixing term\footnote{The reason for implementing the gauge-fixing this way is because obtaining the equation of motion from an action guarantees symmetry of the principal symbol, see section \ref{sec:gaugenull}.}
\begin{equation}
	S_{\rm gauge}=-\frac{1}{2}\int \sqrt{- g} \, H_{a}H^{a} .
\end{equation}
This will contribute to the equations of motion for the metric and the scalar field via terms
\begin{align}
	\frac{1}{\sqrt{-g}} \frac{\delta S_{\rm gauge}}{\delta h_{ab}} & =G^{abcd}\nabla_{c}((1+f)H_{d})\\
	\frac{1}{\sqrt{-g}} \frac{\delta S_{\rm gauge}}{\delta \psi} & = -\nabla_{b}(H^{a} \mathcal{H}_{a}{}^{b})
\end{align}
respectively. We can now write the generalized harmonic gauge linearized equations as
\begin{equation}
	\label{eq:horndeski_lin_ghg}
	\tilde{E}^{(1)}_{ab}=0\qquad \tilde{E}^{(1)}_{\Phi}=0.
\end{equation}
where  
\begin{align}
	\label{eq:horndeski_tilda}
	\tilde{E}^{(1)}_{ab}&=E^{(1)}_{ab}-G_{ab}{}^{cd}\nabla_{c}((1+f)H_{d})\\
	\tilde{E}^{(1)}_{\Phi}&=E^{(1)}_{\Phi}+\nabla_{b}(H^{a} \mathcal{H}_{a}{}^{b}).
\end{align}
It remains to show that the generalized harmonic gauge condition is propagated by the equations of motion. To see this, recall that the action for Horndeski is diffeomorphism invariant, thus for the nonlinear theory we have
\begin{equation}
	0=\int d^4 x \left( \frac{\delta S}{\delta g_{ab}}\nabla_{a}Y_{b}+\frac{\delta S}{\delta \Phi} Y^{b}\nabla_{b}\Phi \right) 
	=\int d^4 x \sqrt{-g}  \left( \nabla^a E_{ab} -E_\Phi \nabla_{b}\Phi\right)Y^{b} .
\end{equation}
This holds for arbitrary $Y^a$ hence, independent of any equation of motion,
\be
  \nabla^a E_{ab} -E_\Phi \nabla_{b}\Phi=0
\ee
and so linearizing around a background solution gives
\begin{equation}
	\nabla^{a}E^{(1)}_{ab}-E^{(1)}_{\Phi}\nabla_{b}\Phi=0.
\end{equation}
Taking the divergence of \eqref{eq:horndeski_tilda} when \eqref{eq:horndeski_lin_ghg} holds and using the above we obtain
\begin{align}
	0&=\nabla^{a}E^{(1)}_{ab}+G_{ab}{}^{cd}\nabla^{b}\nabla_{c}((1+f)H_{d})\nonumber \\
	&=E^{(1)}_{\Phi}\nabla_{b}\Phi-\frac{1}{2}(1+f)(\nabla^{c}\nabla_{c}H_{b}+R_{bc}H^{c})-\nabla^{b}f\nabla_{b}H_{a}-\frac{1}{2}H_{a}\nabla^{b}\nabla_{b} f
\end{align}
that is
\begin{equation}
	(1+f)\nabla_{b}\nabla^{b}H_{a}+2\nabla^{b}f\nabla_{b}H_{a}+2\nabla_{c}(\mathcal{H}^{cd}H_{d})\nabla_{a}\Phi+(1+f)R_{ab}H^{b}+H_{a}\nabla^{b}\nabla_{b}f=0.
\end{equation}
This is a linear wave equation of a standard type for \(H_{a}\), thus, provided that \(H_{a}\) and its time derivative both vanish initially, they will continue to vanish throughout the evolution, i.e. the gauge condition \eqref{eq:ghg} is propagated by the equations of motion \eqref{eq:horndeski_lin_ghg}. It then follows that a solution of the generalized harmonic gauge equations \eqref{eq:horndeski_lin_ghg} is also a solution of the original linearized Horndeski equations of motion \eqref{linhorndeski}.

The linearized generalized harmonic gauge equations of motion \eqref{eq:horndeski_lin_ghg} take the following form 
\begin{align}
	P_{gg}^{abcdef}\nabla_{e}\nabla_{f}h_{cd}+P_{g\Phi}^{abef}\nabla_{e}\nabla_{f}\psi+\ldots = & 0\\
	P_{\Phi g}^{cdef}\nabla_{e}\nabla_{f}h_{cd}+P_{\Phi\Phi}^{ef}\nabla_{e}\nabla_{f}\psi+\ldots = & 0
\end{align}
where the ellipses denotes terms with fewer than $2$ derivatives. We can then define the principal symbol for this system
\begin{equation}
	P(\xi)=\begin{pmatrix}
		P_{gg}^{abcdef}\xi_{e}\xi_{f} & P_{g\Phi}^{abef}\xi_{e}\xi_{f}\\
		P_{\Phi g}^{cdef}\xi_{e}\xi_{f} & P_{\Phi\Phi}^{ef}\xi_{e}\xi_{f}
	\end{pmatrix}
\end{equation}
and we think of it as acting on vectors of the form \((t_{cd},\alpha)^{T}\), where \(t_{cd}\) is a symmetric 2-tensor and \(\alpha\) is a number.

It is convenient to split the principal symbol in its Einstein-scalar-field and Horndeski parts
\begin{equation}
	P(\xi)=P_{\rm Einstein}(\xi)+\delta P(\xi)
\end{equation}
where 
\begin{equation}
	P_{\rm Einstein}(\xi)=
	\begin{pmatrix}
		-\frac{1}{2}\xi^{2}G^{abcd} & 0\\
		0 & -\xi^{2}
	\end{pmatrix}
\end{equation}
is the principal symbol for the harmonic gauge Einstein-scalar-field equations of motion. We write
\be
 \delta P(\xi) = \delta \tilde{P}(\xi) + \delta Q(\xi)
\ee
where $\delta \tilde{P}$ denotes the terms arising from the Horndeski terms $\LL_2, \LL_3,\LL_4, \LL_5$ in the action, and $\delta Q$ denotes the $f$ and $\cal H$-dependent parts of the gauge-fixing terms. Explicitly we have
\begin{equation}
	\delta Q(\xi)=
	\begin{pmatrix}
		-f(f+2) G^{abeh}G_{h}{}^{fcd}\xi_{e}\xi_{f} & (1+f)\xi^{e}G^{fhab}\xi_{h}\mathcal{H}_{ef}\\
		(1+f)\xi^{e}G^{fhcd}\xi_{h}\mathcal{H}_{ef} & -\mathcal{H}_{h}{}^{e}\mathcal{H}^{hf}\xi_{e}\xi_{f}
	\end{pmatrix}.
\end{equation}
From the form of \(P_{\rm Einstein}\) it is clear that all characteristics of the harmonic gauge Einstein-scalar-field system are null. 

We conclude this section by making precise the notion of ``weak background fields'' in the Horndeski setting. We follow a similar approach to the one used for Lovelock theories (cf. Section~\ref{sec:setup}). Consider an orthonormal basis \(\{e_{\mu}\}\) (such that \(e_{0}\) is orthogonal to constant \(t\) surfaces) and denote by \(L^{-2}_{R}\), \(L^{-1}_{1}\) and \(L^{-2}_{2}\) the magnitude of the largest components in such a basis of the Riemann tensor, \(\nabla\Phi\) and \(\nabla\nabla\Phi\) respectively and define \(L^{-2}=\max \{L^{-2}_{R},L^{-2}_{1},L^{-2}_{2}\}\). We want our definition of ``weak fields'' to ensure that the Horndeski terms in the principal symbol are small compared to the Einstein-scalar-field terms, i.e., \(\delta P\) is small compared to \(P_{\rm Einstein}\). This is achieved by requiring the background fields to satisfy
\begin{align}
	\label{eq:horndeski_smallness_2}
	&|\partial^{k}_{X}\GG_{2}|\,L^{-2k+2}\ll1, & k= &1,2, &\\
	\label{eq:horndeski_smallness_3}
	&|\partial^{k}_{X}\partial^{l}_{\Phi}\GG_{3}|\,L^{-2k}\ll1, & k= & 0,1,2, & l= & 0,1, & 1\le k+l\le & 2,\\
	\label{eq:horndeski_smallness_4}
	&|\partial^{k}_{X}\partial^{l}_{\Phi}\GG_{4}|\,L^{-2k}\ll1, & k=& 0,1,2,3, & l=& 0,1,2, & k+l\le & 3,\\
	\label{eq:horndeski_smallness_5}
	&|\partial^{k}_{X}\partial^{l}_{\Phi}\GG_{5}|\,L^{-2k-2}\ll1, & k= & 0,1,2,3, & l= & 0,1,2, & 1\le k+l\le & 3.
\end{align}
We will also require smallness of the functions appearing in the gauge condition:
\be
\label{gauge_small}
|f| \ll 1, \qquad |{\cal H}_\mu{}^\nu| \ll 1 .
\ee
In practice, we will see that strong hyperbolicity will force us to take $f$ and ${\cal H}_a{}^b$ to be particular functions of the background fields, and \eqref{gauge_small} then follows from weakness of the background fields. 

\subsection{Symmetries of the principal symbol}

\label{sec:gaugenull}

For Lovelock theories, our argument for weak hyperbolicity exploited equations (\ref{gaugesym}) following from the identities (\ref{gaugesyma}). Therefore we will need to determine the analogous identities for Horndeski theories. This could be done by explicit computation. Instead we will derive the identities as a consequence of the gauge symmetry of the theory. 
We will appeal to results of Lee and Wald \cite{Lee1990} to do this.

Consider some diffeomorphism covariant theory of gravity, possibly coupled to additional fields, and expand the action to second order around a background solution:
\be
 S = \int d^d x \sqrt{-g} \left( - \frac{1}{2} K^{IJab} \nabla_a u_I \nabla_b u_J + \ldots \right)
\ee
where $u_I$ denotes the perturbation to the fields (including the metric perturbation), the ellipsis denotes terms with fewer than two derivatives, and 
\be
\label{Ksym}
 K^{IJ ab}(x) = K^{JIba}(x).
\ee
Varying the action gives the (linearized) equation of motion
\be
 K^{IJ ab} \nabla_a \nabla_b u_J + \ldots = 0
\ee
where the ellipsis denotes terms with fewer than two derivatives of $u_I$. From this we read off the principal symbol
\be
 P^{IJab} = K^{IJ(ab)}
\ee
so from (\ref{Ksym}) we have
\be
\label{principalsym}
 P^{IJab} = P^{JIab}
\ee
hence symmetry of the principal symbol is a consequence of the variational principle. Varying the action also gives a total derivative term $\nabla_a \theta^a$, where
\be
 \theta^a = -K^{IJab} \delta u_I \nabla_b u_J+ \ldots
\ee
where the ellipsis denotes terms without derivatives. We then define the symplectic current for two independent variations $\delta_1 u_I$ and $\delta_2 u_I$ \cite{Lee1990}
\be
 \omega^a = \delta_1 \theta_2^a - \delta_2 \theta_1^a = K^{IJab} \delta_1 u_I \nabla_b \delta_2 u_J - ( 1 \leftrightarrow 2)+\ldots 
\ee
Given coordinates $(t,x^i)$ where $t$ is a time function, we define the symplectic form as an integral over a surface $\Sigma$ of constant $t$ with unit normal $n_a$
\be
 \omega(\delta_1 u,\delta _2 u) = \int_\Sigma \omega^\mu n_\mu = \int_\Sigma  d^{d-1} x  \sqrt{-g}  \; \omega^0 .
\ee
For a theory with a gauge symmetry, Ref.~\cite{Lee1990} proves that this vanishes if $\delta_2 u$ is taken to be an infinitesimal gauge transformation and $\delta_1 u$ satisfies the (linearized) equation of motion. In particular, it will vanish if $\delta_1 u$ and $\delta_2 u$ are {\it both} infinitesimal gauge transformations. Taking them to be compactly supported gauge transformations we can integrate w.r.t. $t$ to obtain
\be
\label{gaugeint}
 0 = \int d^d x \sqrt{-g} \left[ K^{IJ 0 \nu} \delta_1 u_I \nabla_\nu \delta_2 u_J - ( 1 \leftrightarrow 2) + \ldots  \right].
\ee
As before, the ellipsis denotes terms without derivatives of $\delta_1 u$ or $\delta_2 u$. 

Consider first the case of Lovelock theory (without any gauge-fixing), for which $u_I=h_{ab}$ and we have the symmetries
\be
 K^{abcdef} = K^{bacdef}=K^{abdcef}.
\ee
The gauge transformations are infinitesimal diffeomorphisms:
\be
 \delta h_{ab} = \nabla_{(a} X_{b)}
\ee
where $X_a$ is an arbitrary vector field, assumed compactly supported. Gauge invariance of the action implies, via integration by parts,
\be
0 =  \int d^d x \sqrt{-g}X_b  \left(- K^{abcdef} \nabla_a \nabla_e \nabla_f h_{cd} + \ldots \right)
\ee
where the ellipsis denotes terms with fewer than $3$ derivatives of $h_{\mu\nu}$. Since $X_a$ is arbitrary, this implies
\be
0 = K^{abcdef} \nabla_a \nabla_e \nabla_f h_{cd} + \ldots
\ee
and since $h_{ab}$ is arbitrary, terms with different numbers of derivatives must vanish independently. From the 3-derivative term we obtain
\be
 0 = K^{(a|bcd| ef)} 
\ee
which implies
\be
  P^{(a|bcd|ef)} =0 .
\ee
Now we consider the implications of (\ref{gaugeint}). Take the two gauge transformations to be
\be
 \delta_1 h_{\mu\nu} = \nabla_{(\mu} X_{\nu)} \qquad \delta_2 h_{\mu\nu} = \nabla_{(\mu} Y_{\nu)}
\ee
for arbitrary compactly supported vector fields $X^\mu$, $Y^\mu$. Compact support lets us integrate by parts in (\ref{gaugeint}):
\bea
 0 &=& \int d^d x \sqrt{-g} \left[ \nabla_\mu X_\nu K^{\mu\nu\rho\sigma0 \alpha} \nabla_\alpha \nabla_\rho Y_\sigma -   ( 1 \leftrightarrow 2) + \ldots  \right] \nonumber \\
&=&  \int d^d x \sqrt{-g} X_\nu \left[ -K^{\mu\nu\rho\sigma0 \alpha} \nabla_\mu \nabla_\alpha \nabla_\rho Y_\sigma - K^{\mu\sigma\rho\nu 0 \alpha} \nabla_\alpha \nabla_\rho\nabla_\mu Y_\sigma  + \ldots \right] 
\eea
where the ellipsis denotes terms with fewer than $3$ derivatives of $Y^\mu$. Since $X_\nu$ is arbitrary we must have
\bea
 0 &=& K^{\mu\nu\rho\sigma0 \alpha} \nabla_\mu \nabla_\alpha \nabla_\rho Y_\sigma + K^{\mu\sigma\rho\nu 0 \alpha} \nabla_\alpha \nabla_\rho\nabla_\mu Y_\sigma  + \ldots \nonumber \\
 &=& \left( K^{\mu\nu\rho\sigma0 \alpha}+K^{\mu\sigma\rho\nu 0 \alpha} \right) \partial_\mu \partial_\rho \partial_\alpha Y_\sigma + \ldots \nonumber \\
&=&  \left( K^{\mu\nu\rho\sigma0 \alpha}+K^{\rho\nu \mu\sigma \alpha 0} \right) \partial_\mu \partial_\rho \partial_\alpha Y_\sigma + \ldots 
\eea
Since $Y_\mu$ is arbitrary, the terms with different numbers of derivatives of $Y_\mu$ must vanish independently. Vanishing of the $3$-derivative term requires
\be
0 = K^{\nu(\mu \rho|\sigma0 |\alpha)}+K^{\nu (\rho \mu |\sigma |\alpha) 0} = 2P^{\nu(\mu \rho|\sigma 0| \alpha)}.
\ee
Since the $0$ index refers to an {\it arbitrary} time function $t$, this equation implies
\be
 P^{a(bc|de| f)}=0 .
\ee
The above argument applies to the theory {\it before} fixing the gauge. Of course we can do the same for the Einstein equation. Subtracting the Einstein results from the Lovelock results gives
\be
 \delta P^{(a|bcd| ef)} = \delta P^{a(bc|de| f)} = 0.
\ee
We can now apply this to the harmonic gauge Lovelock equation of motion because the harmonic gauge condition does not affect $\delta P$. In particular we have
\be
\delta P^{abcdef} \xi_a \xi_e \xi_f = \delta P^{abcdef}\xi_b\xi_c  \xi_f  = 0 .
\ee 
Hence we see that the identities (\ref{gaugesym}) are a consequence of the gauge symmetry. 

For a Horndeski theory (before any gauge fixing) we have $u_I = (h_{ab},\psi)$. A gauge transformation is
\be
 \delta h_{ab} = \nabla_{(a} X_{b)} \qquad \delta \psi = X^a \nabla_a \Phi .
\ee
Repeating the above argument for gauge invariance of the action gives
\be
 P_{gg} ^{(a|bcd| ef)} = P_{g\Phi } ^{(a|b | cd)}=0 .
\ee
The symmetry of the principal symbol \eqref{principalsym} then implies that
\be
P_{\Phi g} ^{(a|b | cd)}=P_{g\Phi} ^{(a|b |cd)}=0 .
\ee
Repeating the argument based on (\ref{gaugeint}), the highest derivatives of the gauge transformation parameters $X_\mu$ and $Y_\mu$ arise only from the transformation of $h_{\mu\nu}$ so the result is essentially the same as for Lovelock theory:
\be
 P_{gg} ^{a(bc|de| f)} = 0 .
\ee
These results apply also to the Einstein-scalar-field theory (before gauge fixing). So subtracting the principal symbols for these two cases gives
\be
\label{deltatildePsym}
 0 = \delta \tilde{P}_{gg} ^{(a|bcd|ef)} = \delta {\tilde P}_{g\Phi } ^{(a|b | cd)} = \delta {\tilde P}_{\Phi g } ^{(a|b |cd)}= \delta \tilde{P}_{gg} ^{a(bc|de|f)} .
\ee
Finally, we note that the gauge fixing terms do not affect $\delta \tilde {P}$ so these results apply also to the generalized harmonic gauge equation of motion.  

\subsection{Weak hyperbolicity for weak field background}

We will now begin our study of the hyperbolicity of the linearized Horndeski equations in a generalized harmonic gauge. In this section we will establish weak hyperbolicity of these equations in a weak field background for any generalized harmonic gauge.
Much of the analysis is similar to the analysis of the weak hyperbolicity of harmonic gauge Lovelock theories performed above so we will be briefer here.

As in Section~\ref{sec:setup} we introduce coordinates \(x^{\mu}=(t,x^{i})\) such that \(dt\) is timelike so surfaces of constant $t$ are non-characteristic for the Einstein-scalar-field theory. Again we will denote by \(\xi^{\pm}_{0}\) the two solutions of \(g^{\mu\nu}\xi_{\mu}\xi_{\nu}=0\) for fixed real \(\xi_{i}\), and we define the null covectors $\xi_\mu^\pm = (\xi_0^\pm,\xi_i)$. 

The principal symbol can be regarded as a quadratic form acting on vectors of the form \((t_{\mu\nu},\chi)^{T}\), with \(t_{\mu\nu}\) symmetric. Such vectors form an $11$-dimensional space. Hence \(A,\,B(\xi_{i})\) and \(C(\xi_{i})\) (defined in Sec.~\ref{sec:setup}) are $11 \times 11$ matrices. Explicitly we have
\begin{equation}
	A=\begin{pmatrix}
		P_{gg}^{\mu \nu \rho \sigma 00} & P_{g\Phi}^{\mu \nu 00}\\
		P_{\Phi g}^{\rho \sigma 00}& P_{\Phi \Phi}^{00}
	\end{pmatrix}
	\quad
	B(\xi_{i})=\begin{pmatrix}
		2P_{gg}^{\mu \nu \rho \sigma(0i)}\xi_{i} & 2P_{g\Phi}^{\mu \nu (0i)}\xi_{i}\\
		2P_{\Phi g}^{\rho \sigma (0i)}\xi_{i}&2 P_{\Phi \Phi}^{(0i)}\xi_{i}
	\end{pmatrix}
	\quad
	C(\xi_{i})=\begin{pmatrix}
		P_{gg}^{\mu \nu \rho \sigma ij}\xi_{i}\xi_{j} & P_{g\Phi}^{\mu \nu ij}\xi_{i}\xi_{j}\\
		P_{\Phi g}^{\rho \sigma ij}\xi_{i}\xi_{j}& P_{\Phi \Phi}^{ij}\xi_{i}\xi_{j}
	\end{pmatrix}
\end{equation}
where, again, \(\xi_{i}\) is real and \(\xi_{i}\xi_{i}=1\). These matrices are all real and symmetric: the latter property follows from the fact that the gauge-fixed equations of motion can be derived from an action so \eqref{principalsym} holds. 

For the harmonic gauge Einstein-scalar-field equations, since surfaces of constant $t$ are spacelike, the matrix \(A\) is invertible. By continuity, this will continue to hold for sufficiently weak background fields, once we include the Horndeski terms. Hence we can define real $M(\xi_i)$ as in \eqref{MdefLovelock} and real symmetric $H(\xi_i)$ as in \eqref{HdefLovelock}. These are $22 \times 22$ matrices. As for Lovelock, the matrix $H$ is non-degenerate so its signature can be determined by continuity, i.e., by its signature for the Einstein-scalar-field equations. The result is that it has signature $(11,11)$, i.e., $11$ positive eigenvalues and $11$ negative eigenvalues. As for Lovelock, $M$ is symmetric w.r.t. $H$, i.e., equation (\ref{HMeq}) holds here. 

We consider these matrices as acting on a complex vector space \(V\) of dimension $22$. For the Einstein-scalar-field theory we know that $M$ is diagonalizable with eigenvalues $\xi_0^\pm$, each with degeneracy $11$. So, for linearized Horndeski theory in a weak field background we can proceed as in Sec.~\ref{sec:weakproof} and define the $11$-dimensional subspaces \(V^{\pm}\) as the sum over the generalized eigenspaces of the eigenvectors (of $M$) belonging to the \(\xi^{\pm}_{0}\)-group, respectively. The restriction of $H$ to $V^\pm$ is denoted by $H^\pm$. 

Let us summarize the proof of weak hyperbolicity that we used for Lovelock theories. First we showed that there exist ``pure gauge'' eigenvectors of $M$, with eigenvalue $\xi_0^\pm$. We then showed that such eigenvectors are null and orthogonal w.r.t. $H$ so they form null subspaces $N^\pm$ of $V^\pm$, and that these null subspaces have the maximum dimension consistent with the signature of $H^\pm$. This then excludes the possibility of $M$ possessing a complex eigenvalue $\xi_0$ in, say, the $\xi_0^+$-group, for the corresponding eigenvector would have to be null and orthogonal to $N^+$ so we could add it to $N^+$ to produce a larger null subspace of $V^+$, thereby violating maximality of $N^+$. Hence $M$ cannot have a complex eigenvalue, which establishes weak hyperbolicity. 

All of this extends straightforwardly to Horndeski theories. First note that, as in section \ref{sec:secondorder}, an eigenvector $v$ of $M$ with eigenvalue $\xi_0$ must have the form 
\be
\label{vhorn}
v=\begin{pmatrix} T\\ \xi_{0} T\end{pmatrix}
\ee
where the $11$-vector $T$ must satisfy
\be
\label{PT}
 P(\xi ) T=0
\ee
with $\xi_\mu = (\xi_0,\xi_i)$. We can identify a set of ``pure gauge'' eigenvectors, with eigenvalue $\xi_0^\pm$, given by\footnote{The vanishing of the final component of this vector is related to the fact that under the gauge transformation \eqref{gauge_horn}, the transformation of $\psi$ does not involve a derivative of $Y_a$.} 
\be
\label{Tdef}
 T= \begin{pmatrix} \xi^\pm_{(\mu} X^{}_{\nu)} \\  0 \end{pmatrix}
\ee
for some $X_\mu$. That this satisfies (\ref{PT}) (with $\xi=\xi^\pm$) can be seen as follows. First $P_{\rm Einstein}(\xi^\pm)=0$ because $\xi^\pm_\mu$ is null. Second, the 
results in \eqref{deltatildePsym} imply
\be
\delta \tilde{P}(\xi^\pm) T = 0 .
\ee
Finally, it can be checked explicitly that $\delta Q(\xi^\pm) T = 0$. 

We define $N^\pm$ to be the $4$-dimensional subspace of $V^\pm$ defined by these pure gauge eigenvectors. We now want to prove that $N^\pm$ is null wr.t. $H^\pm$. Consider two pure gauge eigenvectors $v,v' \in N^\pm$ with corresponding $T=(\xi^\pm_{(\mu} X^{}_{\nu)},0)^T$ and $T'=(\xi^\pm_{(\mu} X'^{}_{\nu)},0)^T$. Their inner product w.r.t. $H^\pm$ is the same as their inner product w.r.t. $H$, i.e., as in \eqref{gaugenull}, we have
\be
 v^{'\dagger} H v =  2 \xi_\mu^\pm T^{'\dagger} P^{0\mu} T= 2 \xi_\mu^\pm \xi_\nu^\pm \xi_\rho^\pm \bar{X}'_\sigma X_\tau P_{gg}^{\nu\sigma \rho \tau 0 \mu} =0 
\ee
where the final equality follows from $P_{\rm Einstein}(\xi^\pm)=0$, the final symmetry in \eqref{deltatildePsym}, and the fact that
\be
 \xi_\mu^\pm   \xi_\nu^\pm \xi_\rho^\pm \delta Q_{gg}^{\nu\sigma\rho\tau \lambda \mu}=0 .
\ee
It follows that any two elements of $N^\pm$ are orthogonal w.r.t. $H^\pm$ hence $N^\pm$ defines a $4$-dimensional $H^\pm$-null subspace of $V^\pm$. 

Since $H^\pm$ is the restriction of $H$ to $V^\pm$, it follows that $H^\pm$ is non-degenerate when restricted to $V^\pm$. Hence its signature can be determined by continuity, as we did for Lovelock. In other words, its signature can be determined using the Einstein-scalar-field theory. For this theory, consider two vectors $v_1$ and $v_2$ in $V^\pm$, and hence
of the form (\ref{vhorn}) with $\xi_0 =\xi_0^\pm$. Let the corresponding $11$-vectors be $T_1=(t_{1ab},\chi_1)^T$ and $T_2=(t_{2ab},\chi_2)^T$. The inner product of $v_1$ and $v_2$ w.r.t. $H^\pm$ is the same as the inner product w.r.t. $H$: 
\be
 v_1^\dagger H v_2 = T_1^\dagger B T_2 + 2 \xi_0^\pm T_{1}^{\dagger} A T_2 = 2 \xi_\mu T_1^\dagger P^{0\mu} T_2 = -\xi^{0\pm}\left(  t_1^\dagger G t_2 + \bar{\chi}_1 \chi_2 \right) .
\ee
The argument following \eqref{v1v2} now shows that, when restricted to $V^+$, $H^+$ has $4$ negative eigenvalues and $6+1=7$ positive eigenvalues (the $+1$ coming from $ \bar{\chi}_1 \chi_2$). Similarly for $H^-$ when restricted to $V^-$, with positive and negative interchanged. Hence the dimension of a maximal null subspace of $V^\pm$ is $4$ so $N^\pm$ are maximal null subspaces of $V^\pm$. The proof of weak hyperbolicity follows as explained above. 

\subsection{Strong hyperbolicity of Horndeski theories}

\label{sec:horndeski_strong}

We have shown that, in any generalized harmonic gauge, linearized Horndeski theory is weakly hyperbolic in a weak field background. We will now investigate whether it is also strongly hyperbolic. In particular, strong hyperbolicity requires that $M$ is diagonalizable, i.e., it has no non-trivial Jordan blocks. We can investigate whether or not this is true using the method of Section~\ref{sec:Lovelock_strong}. 

As in section \ref{sec:Lovelock_strong} we define $E^\pm$ to be the generalized eigenspace of the eigenvalue $\xi_0^\pm$. Since $N^\pm \subset E^\pm$ it follows as in section \ref{sec:Lovelock_strong} that $E^\pm$ must have dimension at least $8$.  If $M$ is diagonalizable, then $E^\pm$ are genuine eigenspaces and hence there must exist at least $8$ eigenvectors with eigenvalue $\xi_0^\pm$. So using \eqref{PT} and $P_{\rm Einstein}(\xi^\pm)=0$ we must have $8$ vectors $T$ satisfying $\delta P(\xi^\pm)T=0$. So

\medskip

\noindent {\it A necessary condition for strong hyperbolicity is that, for any null $\xi$, $\ker \delta P(\xi)$ has dimension at least $8$.}

\medskip

Hence strong hyperbolicity implies that, for any null $\xi$, $\ker \delta P(\xi)$ must contain at least $4$ linearly independent ``non-gauge'' elements. 

Let us now look at the condition for a non-trivial Jordan block. As in section \ref{sec:Lovelock_strong}, one can show that the corresponding eigenvalue must be $\xi_0^\pm$ so the block must lie in $E^\pm$. For any such block, there exists a vector $v \in E^\pm$ such that $(M-\xi_0^\pm)v$ is an eigenvector of $M$ with eigenvalue $\xi_0^\pm$ so we must have
\be
\label{2ndorder_horn}
 \left( M - \xi_0^\pm \right) v = \begin{pmatrix} S \\ \xi^\pm_0 S \end{pmatrix} 
\ee
for some non-zero $S=(s_{\mu\nu},\omega)^T$ such that (using $P_{\rm Einstein}(\xi^\pm)=0$)
\be
\label{deltaPS}
 \delta P(\xi^\pm) S = 0 .
\ee
Writing $v=(T,T')^T$ we find that \eqref{2ndorder_horn} reduces to equations analogous to (\ref{2ndorder1}) and \eqref{2ndorder2}:
 \be
  T'=\xi_0^\pm T+ S 
\ee
and 
\be
\label{2ndorder_horn2}
 \delta P(\xi^\pm) T = -(2\xi_0^\pm A + B) S .
\ee
As in section \ref{sec:Lovelock_strong} we contract this with an arbitrary ``pure gauge'' vector $R=(\xi^\pm_{(\mu} X^{}_{\nu)}, 0)^T$. The LHS vanishes and the RHS gives the $H^\pm$-inner product of $(R,\xi_0^\pm R)^T$ with $(S,\xi_0^\pm S)^T$. It follows that $(S,\xi_0^\pm S)^T$ must be $H^\pm$ orthogonal to any pure gauge eigenvector. Similarly, contracting this equation with $S$ and using (\ref{deltaPS}) shows that $(S,\xi_0^\pm S)^T$ is null w.r.t. $H^\pm$. Hence if this vector were not pure gauge then we could add it to $N^\pm$ and violate maximality of this null subspace. Therefore this vector must be pure gauge, i.e., we have $S = (\xi^\pm_{(\mu} Y^{}_{\nu)},0)^T$ for some $Y_\mu \ne 0$. So, writing $T=(t_{\mu\nu},\chi)^T$, \eqref{2ndorder_horn2} takes the form
\begin{equation}
\label{eq:2block}
	\delta P (\xi^{\pm}) \cdot 
	\begin{pmatrix}
		t_{\rho\sigma}\\
		\chi
	\end{pmatrix}= 
	-(2\xi^{\pm}_{0}A+B)\cdot
	\begin{pmatrix}
		\xi^{\pm}_{(\rho}Y^{}_{\sigma)}\\
		0
	\end{pmatrix} .
\end{equation}
If this equation admits a solution for some \(Y_\mu \ne 0 \) then $M$ has a non-trivial Jordan block. So strong hyperbolicity requires that this equation admits no solution $(t_{\mu\nu},\chi)^T$ for any $Y_\mu \ne 0$.

\subsubsection*{Strong hyperbolicity when $\GG_4=\GG_5=0$}

Let us begin by considering the theory with Lagrangian
\begin{equation}
	\LL = \LL_{1}+\LL_{2}+\LL_{3}.
\end{equation}
The nonlinear equations of motion for this theory are
\begin{align}
	E_{ab}  \equiv &\, G_{ab}+\partial_{X}\GG_{3}\left[-\frac{1}{2}\square\Phi \nabla_{a}\Phi\nabla_{b}\Phi+
	 G_{ab}{}^{ed}\nabla^{c}\nabla_{e}\Phi \nabla_{c}\Phi\nabla_{d}\Phi\right]+\ldots=0\\
	E_{\Phi} \equiv &\, -\square\Phi-\partial_{X}\GG_{2}\square\Phi+\partial^{2}_{X}\GG_{2}\nabla^{a}\Phi\nabla^{b}\Phi \nabla_{a}\nabla_{b}\Phi-2\partial_{\Phi}\GG_{3}\square\Phi\nonumber \\
	&-(\partial_{X}\GG_{3}+X\partial^{2}_{X}\GG_{3})\delta^{c_{1}c_{2}}_{d_{1}d_{2}}\nabla_{c_{1}}\nabla^{d_{1}}\Phi\nabla_{c_{2}}\nabla^{d_{2}}\Phi-\frac{1}{2}\partial^{2}_{X}\GG_{3}\delta^{c_{1}c_{2}c_{3}}_{d_{1}d_{2}d_{3}}\nabla_{c_{1}}\nabla^{d_{1}}\Phi\nabla_{c_{2}}\nabla^{d_{2}}\Phi\nabla_{c_{3}}\Phi\nabla^{d_{3}}\Phi\nonumber\\
	&-2\partial^{2}_{X \Phi}\GG_{3}(\delta^{c_{1}c_{2}}_{d_{1}d_{2}}\nabla_{c_{1}}\nabla^{d_{1}}\Phi\nabla_{c_{2}}\Phi\nabla^{d_{2}}\Phi+X\square\Phi)+\partial_{X}\GG_{3} R_{ab}\nabla^{a}\Phi\nabla^{b}\Phi+\ldots=0
\end{align}
where again the ellipsis denotes terms not involving second derivatives. After determining the linearized equations in generalized harmonic gauge \eqref{eq:horndeski_lin_ghg}, we compute the principal symbol and we find that
\begin{align}
	\delta P_{gg}(\xi)^{abcd}= & \, \delta Q_{gg}(\xi)^{abcd}\\
	\delta P_{g \Phi}(\xi)^{ab}=\delta P_{\Phi g}(\xi)^{ab}=& \, -\frac{1}{2}\partial_{X}\GG_{3} \nabla^{a}\Phi\nabla^{b}\Phi \, \xi^{2}+ \xi^{c}G^{deab}\xi_{e}{\cal K}_{cd} \\
	\delta P_{\Phi\Phi}(\xi)=&\, (-\partial_{X}\GG_{2}-2\partial_{\Phi}\GG_{3}+2X\partial^{2}_{X \Phi}\GG_{3}-2\partial_{X}\GG_{3}\square \Phi-2X\partial^{2}_{X}\GG_{3}\square\Phi)\,\xi^{2}\nonumber\\
	&+2\left(\partial_{X}\GG_{3}+X\partial^{2}_{X}\GG_{3}\right)\xi^{c}\xi^{d}\nabla_{c}\nabla_{d}\Phi+(2\partial^{2}_{X\Phi}\GG_{3}+\partial^{2}_{X}\GG_{2})(\xi\cdot\nabla\Phi)^{2}\nonumber\\
	&-\partial^{2}_{X}\GG_{3}\delta^{c_{1}c_{2}c_{3}}_{d_{1}d_{2}d_{3}}\xi_{c_{1}}\xi^{d_{1}}\nabla_{c_{2}}\nabla^{d_{2}}\Phi\nabla_{c_{3}}\Phi\nabla^{d_{3}}\Phi + \delta Q_{\Phi\Phi}(\xi)
\end{align}
where
\be
\label{calKdef}
 {\cal K}_{ab} \equiv (1+f)\mathcal{H}_{ab}+ \partial_{X}\GG_{3}\nabla_{a}\Phi\nabla_{b}\Phi.
\ee
For strong hyperbolicity to hold, equation \eqref{eq:2block} must admit no solution $(t_{\mu\nu},\chi)^T$ when $Y_\mu \ne 0$. Writing out this equation gives
\begin{align}
\label{Yeqa}
	\begin{pmatrix}
		G^{\mu\nu\rho\sigma}[-f(f+2)G_{\rho}{}^{\lambda\alpha\beta}\xi^{\pm}_{\sigma}\xi^{\pm}_{\lambda}t_{\alpha\beta}+\xi^{\pm \lambda}{\cal K}_{\lambda\rho}\xi^{\pm}_{\sigma}\chi] \\ 
		\\
		\xi^{\pm \mu}G^{\nu \lambda\rho\sigma}\xi^{\pm}_{\lambda}t_{\rho\sigma}{\cal K}_{\mu\nu}+\delta P_{\Phi \Phi}(\xi^{\pm})\chi
	\end{pmatrix}=
	\begin{pmatrix}
		\xi^{0 \pm}G^{\mu\nu\rho\sigma}\xi^{\pm}_{\rho}Y^{}_{\sigma}\\
		\\
		(\partial_{X} \GG_{3})\xi^{0 \pm} (\xi^{\pm}\cdot\nabla\Phi)(Y\cdot \nabla\Phi)-\mathcal{K}_{\lambda\sigma}\xi^{\pm \lambda}G^{\mu\nu\sigma 0}\xi^{\pm}_{\mu}Y^{}_{\nu}
	\end{pmatrix}.
\end{align}
Looking at the first row of this equation, the non-degeneracy of \(G^{\mu\nu\rho\sigma}\) implies that if $f \ne 0$ then we can solve for the ``non-transverse'' part of $t_{\mu\nu}$:\footnote{Note that our smallness assumption \eqref{gauge_small} implies that $f \ne -2$.}
\be
 G_{\mu}{}^{\nu\rho\sigma} \xi_{\nu}^{\pm} t_{\rho\sigma} = \frac{1}{f(f+2)}(\xi^{\pm \rho}\mathcal{K}_{\rho \mu}\chi-\xi^{0\pm}Y_{\mu}) .
\ee
This can then be substituted into the second row to obtain an equation that determines $\chi$. Hence if $f\ne 0$ then a solution of \eqref{eq:2block} exists for any $Y_\mu \ne 0$. Therefore strong hyperbolicity requires that $f=0$. With $f=0$, the first row of \eqref{Yeqa} implies
\be
\label{Ysol}
 \xi^{0\pm} Y_\mu = \xi^{\pm \rho}\mathcal{K}_{\rho\mu}\chi .
\ee 
Plugging this into the second row of \eqref{Yeqa} now gives a linear homogeneous scalar equation for $\chi$ and $t_{\mu\nu}$. Since this is only one equation for $11$ unknowns, there exist non-trivial solutions. We see that we can solve \eqref{eq:2block} for $Y_\mu$ of the form \eqref{Ysol}. Hence if this $Y_\mu$ is non-vanishing then the equation is not strongly hyperbolic. Therefore strong hyperbolicity, requires \eqref{Ysol} to vanish for any (null) $\xi_\mu^\pm$ which implies (since generically $\chi \ne 0$) ${\cal K}_{\mu\nu}=0$. Hence strong hyperbolicity selects a unique generalized harmonic gauge:
\be
\label{eq:gengauge_3}
f=0 \qquad \mathcal{H}_{ab}= -\partial_{X}\GG_{3}\nabla_{a}\Phi\nabla_{b}\Phi.
\ee
Note that this guarantees that the smallness condition \eqref{gauge_small} is satisfied. If our gauge functions $f$ and ${\cal H}_{ab}$ satisfy this equation then $M$ is diagonalizable, as required by strong hyperbolicity. As explained above \eqref{Hcond}, diagonalizability ensures that there exists a positive definite symmetrizer $K$ satisfying \eqref{Hcond2}. To complete the proof of strong hyperbolicity we need to check that $K$ depends smoothly on $\xi_i$. We will do this in a more general setting below (see the discussion below Eq.~\eqref{eq:gengauge_4_phi}).

\subsubsection*{Failure of strong hyperbolicity when $\partial_X \GG_4 \ne 0$, $\GG_5=0$}

The situation is different if we include \(\LL_{4}\) i.e. we work with the theory
\begin{equation}
\label{quartic}
	\LL = \LL_{1}+\LL_{2}+\LL_{3}+\LL_{4}
\end{equation} 
We will show that if $\partial_X \GG_4 \ne 0$ then there is no generalized hyperbolic gauge for which this theory is strongly hyperbolic.

The terms in the equations of motion $E^a{}_b$ and $E_\Phi$ arising from $\LL_4$ are \cite{Kobayashi2011,Gao2011} 
\begin{align}
	E^a{}_b^{(4)}=& \,(\GG_{4}-2X\partial_{X}\GG_{4})G^{a}{}_{b}+\frac{1}{4}\partial_{X}\GG_{4}\delta^{ac_{1}c_{2}c_{3}}_{bd_{1}d_{2}d_{3}}R_{c_{1}c_{2}}{}^{d_{1}d_{2}}\nabla_{c_{3}}\Phi\nabla^{d_{3}}\Phi\nonumber\\
	& +\frac{1}{2}(\partial_{X}\GG_{4}+2X\partial^{2}_{X}\GG_{4})\delta^{a c_{1} c_{2}}_{b d_{1} d_{2}}\nabla_{c_{1}}\nabla^{d_{1}}\Phi\nabla_{c_{2}}\nabla^{d_{2}}\Phi\nonumber\\
	& +\frac{1}{2}\partial^{2}_{X}\GG_{4}\delta^{a c_{1} c_{2} c_{3}}_{b d_{1} d_{2} d_{3}}\nabla_{c_{1}}\nabla^{d_{1}}\Phi\nabla_{c_{2}}\nabla^{d_{2}}\Phi\nabla_{c_{3}}\Phi\nabla^{d_{3}}\Phi\nonumber\\
	& + 2\partial^{2}_{X\Phi}\GG_{4}\delta^{a c_{1} c_{2}}_{b d_{1} d_{2}}\nabla_{c_{1}}\nabla^{d_{1}}\Phi \nabla_{c_{2}}\Phi \nabla^{d_{2}}\Phi+(\partial_{\Phi}\GG_{4}+2X\partial^{2}_{X\Phi}\GG_{4})\delta^{a c_{1}}_{b d_{1}}\nabla_{c_{1}}\nabla^{d_{1}}\Phi\\
	E_\Phi^{(4)}=& \,-\frac{1}{2}(\partial_{X}\GG_{4}+2X\partial^{2}_{X}\GG_{4})\delta^{c_{1}c_{2}c_{3}}_{d_{1}d_{2}d_{3}}\nabla_{c_{1}}\nabla^{d_{1}}\Phi R_{c_{2}c_{3}}{}^{d_{2}d_{3}}-(\partial_{\Phi}\GG_{4}+2X\partial^{2}_{X\Phi}\GG_{4})R \nonumber\\
	& -\frac{1}{2}\partial^{2}_{X}\GG_{4}\delta^{c_{1}c_{2}c_{3}c_{4}}_{d_{1}d_{2}d_{3}d_{4}}\nabla_{c_{1}}\nabla^{d_{1}}\Phi\nabla_{c_{2}}\Phi\nabla^{d_{2}}\Phi R_{c_{3}c_{4}}{}^{d_{3}d_{4}} \nonumber\\
	& -\partial^{2}_{X\Phi}\GG_{4}\delta^{c_{1}c_{2}c_{3}}_{d_{1}d_{2}d_{3}}\nabla_{c_{1}}\Phi\nabla^{d_{1}}\Phi R_{c_{2}c_{3}}{}^{d_{2}d_{3}}\nonumber \\
	& - \left(\partial^{2}_{X}\GG_{4}+\frac{2}{3}X\partial^{3}_{X}\GG_{4}\right)\delta^{c_{1}c_{2}c_{3}}_{d_{1}d_{2}d_{3}}\nabla_{c_{1}}\nabla^{d_{1}}\Phi\nabla_{c_{2}}\nabla^{d_{2}}\Phi\nabla_{c_{3}}\nabla^{d_{3}}\Phi \nonumber \\
	& -2\partial^{3}_{X\Phi\Phi}\GG_{4}\delta^{c_{1}c_{2}}_{d_{1}d_{2}}\nabla_{c_{1}}\Phi\nabla^{d_{1}}\Phi\nabla_{c_{2}}\nabla^{d_{2}}\Phi \nonumber \\
	& -(2X\partial^{3}_{X X\Phi}\GG_{4}+3\partial^{2}_{X\Phi}\GG_{4})\delta^{c_{1}c_{2}}_{d_{1}d_{2}}\nabla_{c_{1}}\nabla^{d_{1}}\Phi\nabla_{c_{2}}\nabla^{d_{2}}\Phi \nonumber \\
	& - 2\partial^{3}_{XX\Phi}\GG_{4}\delta^{c_{1}c_{2}c_{3}}_{d_{1}d_{2}d_{3}}\nabla_{c_{1}}\nabla^{d_{1}}\Phi\nabla_{c_{2}}\nabla^{d_{2}}\Phi\nabla_{c_{3}}\Phi\nabla^{d_{3}}\Phi \nonumber \\
	& -\frac{1}{3}\partial^{3}_{X}\GG_{4}\delta^{c_{1}c_{2}c_{3}c_{4}}_{d_{1}d_{2}d_{3}d_{4}}\nabla_{c_{1}}\nabla^{d_{1}}\Phi\nabla_{c_{2}}\nabla^{d_{2}}\Phi\nabla_{c_{3}}\nabla^{d_{3}}\Phi\nabla_{c_{4}}\Phi\nabla^{d_{4}}\Phi.
\end{align}
Linearizing these equations, and including the gauge-fixing terms, one can then compute \(\delta \tilde{P}^{(4)}\), the contribution to $\delta \tilde{P}$ arising from \(\LL_{4}\). It takes the following form
\begin{align}
	\delta \tilde{P}^{(4)}_{gg} (\xi)^{a}{}_{b}{}^{cd}t_{cd}=& -\frac{1}{2}(\GG_{4}-2X\partial_{X}\GG_{4})\delta^{a c_{1}c_{2}}_{bd_{1} d_{2}}\xi_{c_{1}}\xi^{d_{1}} t_{c_{2}}{}^{d_{2}} \nonumber \\
	 &-\frac{1}{2}\partial_{X}\GG_{4} \delta^{a c_{1} c_{2} c_{3}}_{b d_{1} d_{2} d_{3}}\xi_{c_{1}}\xi^{d_{1}}t_{c_{2}}{}^{d_{2}}\nabla_{c_{3}}\Phi \nabla^{d_{3}} \Phi \\
	\delta \tilde{P}^{(4)}_{g\Phi} (\xi)^{a}{}_{b}=\delta \tilde{P}^{(4)}_{\Phi g} (\xi)^{a}{}_{b}=& (\partial_{X}\GG_{4}+2X \partial^{2}_{X}\GG_{4})\delta^{a c_{1}c_{2}}_{b d_{1}d_{2}} \xi_{c_{1}}\xi^{d_{1}}\nabla_{c_{2}}\nabla^{d_{2}}\Phi \nonumber\\
	&+\partial^{2}_{X}\GG_{4}\delta^{a c_{1}c_{2}c_{3}}_{b d_{1} d_{2} d_{3}} \xi_{c_{1}}\xi^{d_{1}}\nabla_{c_{2}}\nabla^{d_{2}}\Phi \nabla_{c_{3}}\Phi\nabla^{d_{3}}\Phi \nonumber\\
	&+2\partial^{2}_{X\Phi}\GG_{4}\delta^{a c_{1}c_{2}}_{b d_{1} d_{2}}\xi_{c_{1}}\xi^{d_{1}}\nabla_{c_{2}}\Phi\nabla^{d_{2}}\Phi \nonumber \\
	&+(\partial_{\Phi}\GG_{4}+2X\partial^{2}_{X\Phi}\GG_{4})\delta^{a c_{1}}_{b d_{1}}\xi_{c_{1}}\xi^{d_{1}}\\
	\delta \tilde{P}^{(4)}_{\Phi\Phi} (\xi)=&- \frac{1}{2}(\partial_{X}\GG_{4}+2X\partial^{2}_{X}\GG_{4})\delta^{c_{1}c_{2}c_{3}}_{d_{1}d_{2}d_{3}}\xi_{c_{1}}\xi^{ d_{1}}R_{c_{2}c_{3}}{}^{d_{2}d_{3}} \nonumber\\
	&-\frac{1}{2}\partial^{2}_{X}\GG_{4}\delta^{c_{1}c_{2}c_{3}c_{4}}_{d_{1}d_{2}d_{3}d_{4}}\xi_{c_{1}}\xi^{d_{1}}\nabla_{c_{2}}\Phi\nabla^{d_{2}}\Phi R_{c_{3} c_{4}}{}^{d_{3}d_{4}} \nonumber\\
	&-(3\partial^{2}_{X}\GG_{4}+2 X \partial^{3}_X\GG_{4})\delta^{c_{1}c_{2}c_{3}}_{d_{1}d_{2}d_{3}}\xi_{c_{1}}\xi^{d_{1}}\nabla_{c_{2}}\nabla^{d_{2}}\Phi \nabla_{c_{3}}\nabla^{d_{3}}\Phi \nonumber \\
	&-\partial^{3}_{X}\GG_{4}\delta^{c_{1}c_{2}c_{3}c_{4}}_{d_{1}d_{2}d_{3}d_{4}}\xi_{c_{1}}\xi^{d_{1}}\nabla_{c_{2}}\nabla^{d_{2}}\Phi \nabla_{c_{3}}\nabla^{d_{3}}\Phi\nabla_{c_{4}}\Phi \nabla^{d_{4}}\Phi \nonumber\\
	&-4\partial^{3}_{XX\Phi}\GG_{4}\delta^{c_{1}c_{2}c_{3}}_{d_{1}d_{2}d_{3}}\xi_{c_{1}}\xi^{d_{1}}\nabla_{c_{2}}\nabla^{d_{2}}\Phi \nabla_{c_{3}}\Phi \nabla^{d_{3}}\Phi\nonumber \\
	&-2(2X\partial^{3}_{X X\Phi}\GG_{4}+3\partial^{2}_{X\Phi}\GG_{4})\delta^{c_{1}c_{2}}_{d_{1}d_{2}}\xi_{c_{1}}\xi^{ d_{1}}\nabla_{c_{2}}\nabla^{d_{2}}\Phi\nonumber\\
	&-2\partial^{3}_{X\Phi\Phi}\GG_{4}\delta^{c_{1}c_{2}}_{d_{1}d_{2}}\xi_{c_{1}}\xi^{ d_{1}}\nabla_{c_{2}}\Phi\nabla^{d_{2}}\Phi.	
\end{align}
As discussd above, for the equations to be strongly hyperbolic it is necessary that the kernel of \(\delta P (\xi^{\pm})\) has dimension $8$ or greater. We will now study whether this condition is satisfied. A vector \((t_{ab},\chi)^{T}\) is in \(\ker \delta P (\xi^{\pm})\) if, and only if,
\begin{equation}
	\label{eq:kernel_dP4}
	\begin{pmatrix}
		\delta P_{gg}(\xi^{\pm})^{abcd}t_{cd}+\delta P_{g\Phi}(\xi^{\pm})^{ab}\chi\\
		\delta P_{\Phi g}(\xi^{\pm})^{cd}t_{cd}+\delta P_{\Phi\Phi}(\xi^{\pm})\chi
	\end{pmatrix}
	=0 .
\end{equation}
We now assume that \(\partial_{X}\GG_{4}\neq 0\). In this case we can separate out a term proportional to \(t_{ab}\) in the first row of \eqref{eq:kernel_dP4} and write this equation as
\begin{align}
	(\xi^{\pm}\cdot \nabla\Phi)^{2}t_{ab}=&
	-\xi^{\pm}_{a}\xi^{\pm}_{b}(\nabla^{c}\Phi\nabla^{d}\Phi t_{cd}+\GG_{4} t^{c}{}_{c})\nonumber\\
	&+2\xi^{\pm}_{(a}t^{}_{b)c}\left(\frac{\GG_{4}}{\partial_{X}\GG_{4}} \xi^{\pm c}+\nabla^{c}\Phi (\xi^{\pm}\cdot \nabla\Phi)\right)
	-2\xi^{\pm}_{(a}\nabla^{}_{b)}\Phi (t^{c}{}_{c}(\xi^{\pm}\cdot\nabla\Phi)-\xi^{\pm c}\nabla^{d}\Phi t_{cd})\nonumber\\
	&-g_{ab}\left(2\xi^{\pm c}\nabla^{d}\Phi t_{cd}(\xi^{\pm}\cdot \nabla\Phi)+\frac{\GG_{4}}{\partial_{X}\GG_{4}}\xi^{\pm c}\xi^{\pm d}t_{cd}-t^{c}{}_{c}(\xi^{\pm}\cdot\nabla\Phi)^{2}\right)\nonumber\\
	&-\nabla_{(a}\Phi\nabla_{b)}\Phi \xi^{\pm c}\xi^{\pm d}t_{cd}-\nabla_{(a}\Phi t_{b)c}\xi^{\pm c} (\xi^{\pm}\cdot\nabla\Phi)\nonumber\\
	&+\frac{2}{\partial_{X}\GG_{4}}(\delta Q_{gg}(\xi^{\pm})_{ab}{}^{cd}t_{cd}+\delta P_{g\Phi}(\xi^{\pm})_{ab} \chi).
\end{align}
Note that for a generic weak-field background, and generic $\xi^\pm$, we have $\xi^\pm \cdot \nabla \Phi \ne 0$. 
From the tensor structure of this equation, we deduce that \(t_{ab}\) must take the form
\begin{equation}
\label{t_ans}
	t_{ab}= \xi^{\pm}_{(a}Y^{}_{b)}+\lambda g_{ab}+Z_{(a}\nabla_{b)}\Phi+\mu \nabla_{a}\nabla_{b}\Phi
\end{equation}
for some $Y_a$, $\lambda$, $Z_a$ and $\mu$. The last term in this expression comes from the fact that \(\delta P_{g\Phi}(\xi^{\pm})_{ab}\) contains terms proportional to \(\nabla_{a}\nabla_{b}\Phi\) as well as terms of the other three types. There is some degeneracy in this expression e.g. degeneracy between the first and third terms implies that $Z_a$ is defined only up to addition of a multiple of $\xi_a^\pm$, i.e., the part of $Z_a$ parallel to $\xi_a^\pm$ is ``pure gauge''. For strong hyperbolicity we need there to exist at least $4$ linearly independent ``non-gauge'' elements of \(\ker \delta P(\xi^{\pm})\). The first term in \eqref{t_ans} is pure gauge. The ``non-gauge part'' is determined by $\chi,\lambda,\mu$ and the non-gauge part of $Z_a$.   

Plugging \eqref{t_ans} back into the first row of \eqref{eq:kernel_dP4} we get
\begin{align}
	\label{eq:kernel_dP4_g}
	0=\delta P_{gg}(\xi^{\pm})^{a}{}_{b}{}^{cd}t_{cd}+\delta P_{g\Phi}(\xi^{\pm})^{a}{}_{b}\chi=&\,
	\delta^{a c_{1} c_{2} c_{3}}_{b d_{1} d_{2} d_{3}}\xi^{\pm}_{c_{1}}\xi^{\pm d_{1}}\nabla_{c_{2}}\nabla^{d_{2}}\Phi \nabla_{c_{3}}\Phi\nabla^{d_{3}}\Phi \left(-\frac{1}{2}\partial_{X}\GG_{4} \mu + \partial^{2}_{X}\GG_{4}\chi\right) \nonumber\\
	&+\delta^{a c_{1} c_{2}}_{b d_{1} d_{2}}\xi^{\pm}_{c_{1}}\xi^{\pm d_{1}}\biggl[\nabla_{c_{2}}\Phi\nabla^{d_{2}}\Phi
	\left(-\frac{1}{2}\partial_{X}\GG_{4}\lambda+2\partial^{2}_{X\Phi}\GG_{4}\chi\right) \nonumber\\
	&-\frac{1}{4}(\GG_{4}-2X\partial_{X}\GG_{4}-f(f+2))(Z_{c_{2}}\nabla^{d_{2}}\Phi+\nabla_{c_{2}}\Phi Z^{d_{2}})\biggr]\nonumber\\
	&+\delta^{a c_{1} c_{2}}_{b d_{1} d_{2}}\xi^{\pm}_{c_{1}}\xi^{\pm d_{1}}\nabla_{c_{2}}\nabla^{d_{2}}\Phi\biggl[(\partial_{X}\GG_{4}+2X\partial^{2}_{X}\GG_{4})\chi\nonumber\\
	& -\frac{1}{2}\mu(\GG_{4}-2X\partial_{X}\GG_{4}-f(f+2))\biggr] \nonumber\\
	&-\biggl[(\partial_{\Phi}\GG_{4}+2 X\partial^{2}_{X\Phi}\GG_{4})\chi-\lambda(\GG_{4}-2X\partial_{X}\GG_{4}-f(f+2))\biggr]\xi^{\pm a} \xi^{\pm}_{b}\nonumber\\
	&+\xi^{\pm c}G^{dea}{}_{b}\xi^{\pm}_{e}\mathcal{K}_{cd}\chi
\end{align}
where ${\cal K}_{ab}$ is defined in \eqref{calKdef}. We will now show that the requirement of strong hyperbolicity fixes our choice of gauge. Consider first the case
\begin{equation}
\label{Z_cond}
	\GG_{4}-2X\partial_{X}\GG_{4}-f(f+2)\neq 0 .
 \end{equation}
In this case, \eqref{eq:kernel_dP4_g} contains $Z_a$-dependent terms proportional to
\begin{equation}
	\delta^{a c_{1} c_{2}}_{b d_{1} d_{2}}\xi^{\pm}_{c_{1}}\xi^{\pm d_{1}}(Z_{c_{2}}\nabla^{d_{2}}\Phi+\nabla_{c_{2}}\Phi Z^{d_{2}})=4G^{a}{}_{b}{}^{ce}\xi^{\pm d}\xi^{\pm}_{e}G_{cd}{}^{fh}Z_{f}\nabla_{h}\Phi.
\end{equation}
View the RHS as an operator ${\cal O}$ acting on $Z_a$. Let's determine the kernel of this operator. Since \(G^{abcd}\) is non-degenerate, vectors in the kernel must satisfy
\be
 \label{eq:kernel_O}
 \xi^{\pm d}\xi^{\pm}_{(e}G_{c)d}{}^{fh}Z_{f}\nabla_{h}\Phi =0\qquad \Rightarrow \qquad \xi^{\pm d}G_{cd}{}^{fh}Z_{f}\nabla_{h}\Phi =0 .
\ee
However, for generic $\nabla_a \Phi$, it is easy to show that all solutions of this equation have $Z_a$ proportional to $\xi^\pm_a$. Hence the kernel of ${\cal O}$ generically contains only vectors proportional to $\xi^\pm_a$. This implies that, generically, if equation \eqref{eq:kernel_dP4_g} admits a solution then $Z_a$ is determined up to a multiple of $\xi^\pm_a$, in terms of $\chi,\lambda,\mu$. In other words, the non-gauge part of $Z_a$ is fixed uniquely by the $3$ quantities $\chi,\lambda,\mu$. Therefore, there exist at most $3$ linearly independent non-gauge elements of \(\ker \delta P(\xi^{\pm})\), whereas strong hyperbolicity requires at least $4$ such elements. So if our gauge condition satisfies \eqref{Z_cond} then the equation is not strongly hyperbolic. 

We have shown that strong hyperbolicity requires that our gauge function $f$ obeys
\begin{equation}
	\GG_{4}-2X\partial_{X}\GG_{4}-f(f+2)= 0.
\end{equation}
We can solve this quadratic equation and choose the root that satisfies the smallness condition \eqref{gauge_small} when the conditions \eqref{eq:horndeski_smallness_4} are satisfied:
\begin{equation}
	f=-1+\sqrt{1+\GG_{4}-2X\partial_{X}\GG_{4}} .
\end{equation}
The contraction of \eqref{eq:kernel_dP4_g} with \(\nabla^b\Phi\) gives
\be
	\label{eq:kernel_dP4_g_contracted}
	0=\xi^{\pm c} \xi^{\pm}_{e} \nabla_{b}\Phi G^{deab}\tilde{\cal K}_{cd} \chi
\ee
where
\begin{equation}
	\tilde{\cal K}_{cd}\equiv \mathcal{K}_{cd}-(\alpha g_{cd}+\beta \nabla_{c}\nabla_{d}\Phi)
\end{equation}
with
\begin{equation}
	\alpha=\partial_{\Phi}\GG_{4}+2X\partial^{2}_{X\Phi}\GG_{4}+\nabla^{a}\nabla_{a}\Phi (\partial_{X}\GG_{4}+2X\partial^{2}_{X}\GG_{4})\qquad \beta=-2(\partial_{X}\GG_{4}+2X\partial^{2}_{X}\GG_{4}).
\end{equation}
Consider first the case in which our gauge condition is such that, generically,  
\be
\xi^{\pm c} \xi^{\pm}_{e} \nabla_{b}\Phi G^{deab}\tilde{\cal K}_{cd} \ne 0 .
\ee
Then, in a generic background, for generic null $\xi_a^\pm$, \eqref{eq:kernel_dP4_g_contracted} implies that we must have $\chi=0$ and equation \eqref{eq:kernel_dP4_g} then reduces to
\be
   0=
	-\frac{1}{2}\partial_{X}\GG_{4} \mu \,\delta^{a c_{1} c_{2} c_{3}}_{b d_{1} d_{2} d_{3}}\xi^{\pm}_{c_{1}}\xi^{\pm d_{1}}\nabla_{c_{2}}\nabla^{d_{2}}\Phi \nabla_{c_{3}}\Phi\nabla^{d_{3}}\Phi 
	 -\partial_{X}\GG_{4}\lambda \,G^{a}{}_{b}{}^{ec}\xi^{\pm d}\xi^{\pm}_{e}G_{cd}{}^{fh}\nabla_{f}\Phi\nabla_{h}\Phi .
\ee
In a generic background this implies \(\lambda=\mu=0\) (using $\partial_X \GG_4 \ne 0$). But with $\chi=\lambda=\mu=0$, the ``non-gauge'' part of the vector $(t_{ab},\chi)^T$ is determined entirely by $Z_a$ which has at most $3$ independent non-gauge components. So in this case we do not have enough non-gauge elements of \(\ker \delta P(\xi^{\pm})\) for strong hyperbolicity.

We have shown that strong hyperbolicity requires that, generically,
\be
\xi^{\pm c} \xi^{\pm}_{e} \nabla_{b}\Phi G^{deab}\tilde{\cal K}_{cd} = 0 .
\ee
For this to be satisfied for generic null \(\xi^{\pm}\) we must have
\begin{equation}
	\label{eq:fix_K_tilde_L4}
	G^{abde}\nabla_{b}\Phi\tilde{\cal K}_{cd}=\rho^{a}\delta^{e}_{c}
\end{equation}
for some vector \(\rho^{a}\). Contracting with \(\nabla_{a}\Phi\) we see that
\begin{equation}
	(\nabla\Phi)^{2}\tilde{\cal K}_{a}{}^{b}=2\tilde{\cal K}_{ac}\nabla^{c}\Phi\nabla^{b}\Phi-2(\rho\cdot\nabla\Phi)\delta^{b}_{a}
\end{equation}
from which we deduce that the most general form \(\tilde{\cal K}\) can take is
\begin{equation}
	\tilde{\cal K}_{ab}=\kappa g_{ab}+W_{a}\nabla_{b}\Phi.
\end{equation}
for some scalar \(\kappa\) and vector \(W_{a}\). Note that we can determine \(\rho^{a}\) in terms of these quantities by taking the trace over the \(e\) and \(c\) indices in \eqref{eq:fix_K_tilde_L4}
\begin{equation}
	\rho^{a}=\frac{1}{4}\left(-\kappa \nabla^{a}\Phi+G^{abcd}W_{c}\nabla_{d}\Phi\nabla_{b}\Phi\right).
\end{equation}
Plugging these back into \eqref{eq:fix_K_tilde_L4} we find that the only solution is given by \(\kappa=0\) and \(W_{a}=0\), that is 
\begin{equation}
	\label{eq:fix_K_L4}
	\tilde{\cal K}_{ab}=0.
\end{equation}
Hence strong hyperbolicity for a generic weak-field background forces us to make the gauge choice
\begin{align}
	\label{eq:gengauge_4}
	f&=-1+\sqrt{1+\GG_{4}-2X\partial_{X}\GG_{4}}\nonumber\\
	(1+f)\mathcal{H}_{ab}&=\alpha g_{ab}+ \beta \nabla_{a}\nabla_{b}\Phi - \partial_{X}\GG_{3} \nabla_{a}\Phi\nabla_{b}\Phi.
\end{align}
With this choice of gauge, \eqref{eq:kernel_dP4_g} reduces to
\begin{align}
    0=&\,
	\delta^{a c_{1} c_{2} c_{3}}_{b d_{1} d_{2} d_{3}}\xi^{\pm}_{c_{1}}\xi^{\pm d_{1}}\nabla_{c_{2}}\nabla^{d_{2}}\Phi \nabla_{c_{3}}\Phi\nabla^{d_{3}}\Phi \left(-\frac{1}{2}\partial_{X}\GG_{4} \mu + \partial^{2}_{X}\GG_{4}\chi\right) \nonumber\\
	&+2G^{a}{}_{b}{}^{ec}\xi^{\pm d}\xi^{\pm}_{e}G_{cd}{}^{fh}\nabla_{f}\Phi\nabla_{h}\Phi
	\left(-\frac{1}{2}\partial_{X}\GG_{4}\lambda+2\partial^{2}_{X\Phi}\GG_{4}\chi\right) .
\end{align}
For a generic background, this fixes \(\lambda\) and \(\mu\) in terms of \(\chi\):
\begin{equation}
	\lambda=4\frac{\partial^{2}_{X\Phi}\GG_{4}}{\partial_{X}\GG_{4}}\chi 
	\qquad \mu=2\frac{\partial^{2}_{X}\GG_{4}}{\partial_{X}\GG_{4}}\chi.
\end{equation}
We now consider the second row of \eqref{eq:kernel_dP4}, which takes the form 
\be
\mathbb{A} \chi=0
\ee  
where
\begin{align}
	\mathbb{A}=&- \frac{1}{2}[\partial_{X}\GG_{4}+2X\partial^{2}_{X}\GG_{4}]\delta^{c_{1}c_{2}c_{3}}_{d_{1}d_{2}d_{3}}\xi^{\pm}_{c_{1}}\xi^{\pm d_{1}}R_{c_{2}c_{3}}{}^{d_{2}d_{3}} \nonumber\\
	&-\frac{1}{2}\partial^{2}_{X}\GG_{4}\delta^{c_{1}c_{2}c_{3}c_{4}}_{d_{1}d_{2}d_{3}d_{4}}\xi^{\pm}_{c_{1}}\xi^{\pm d_{1}}\nabla_{c_{2}}\Phi\nabla^{d_{2}}\Phi R_{c_{3} c_{4}}{}^{d_{3}d_{4}} \nonumber\\
	&-\biggl[3\partial^{2}_{X}\GG_{4}+2 X \partial^{3}_X\GG_{4}+\frac{2(\partial_{X}\GG_{4}+2X \partial^{2}_{X}\GG_{4})^{2}}{1+\GG_{4}-2X\partial_{X}\GG_{4}}\biggr]\delta^{c_{1}c_{2}c_{3}}_{d_{1}d_{2}d_{3}}\xi^{\pm}_{c_{1}}\xi^{\pm d_{1}}\nabla_{c_{2}}\nabla^{d_{2}}\Phi \nabla_{c_{3}}\nabla^{d_{3}}\Phi \nonumber \\
	&+[2(\partial_{X}\GG_{4})^{-1}(\partial^{2}_{X}\GG_{4})^{2}-\partial^{3}_{X}\GG_{4}]\delta^{c_{1}c_{2}c_{3}c_{4}}_{d_{1}d_{2}d_{3}d_{4}}\xi^{\pm}_{c_{1}}\xi^{\pm d_{1}}\nabla_{c_{2}}\nabla^{d_{2}}\Phi \nabla_{c_{3}}\nabla^{d_{3}}\Phi\nabla_{c_{4}}\Phi \nabla^{d_{4}}\Phi \nonumber\\
	&+4\biggl[\partial^{3}_{XX\Phi}\GG_{4}+2\frac{\partial^{2}_{X}\GG_{4}\partial^{2}_{X \Phi}\GG_{4}}{\partial_{X}\GG_{4}}-\frac{\partial_{X}\GG_{3}(\partial_{X}\GG_{4}+2X\partial^{2}_{X}\GG_{4})}{2(1+\GG_{4}-2X\partial_{X}\GG_{4})}-\partial^{2}_{X}\GG_{3}\biggr]\delta^{c_{1}c_{2}c_{3}}_{d_{1}d_{2}d_{3}}\xi^{\pm}_{c_{1}}\xi^{\pm d_{1}}\nabla_{c_{2}}\nabla^{d_{2}}\Phi \nabla_{c_{3}}\Phi \nabla^{d_{3}}\Phi\nonumber \\
	&+2\biggl[-8\frac{( \partial^{2}_{X\Phi}\GG_{4})^{2}}{\partial_{X}\GG_{4}}+\frac{\partial_{X}\GG_{3}(\partial_{\Phi}\GG_{4}+2X\partial^{2}_{X\Phi}\GG_{4}+X\partial_{X}\GG_{3})}{(1+\GG_{4}-2X\partial_{X}\GG_{4})}+\left(\partial^{2}_{X\Phi}\GG_{3}+\frac{1}{2}\partial^{2}_{X}\GG_{2}+\partial^{3}_{X\Phi\Phi}\GG_{4}\right)\biggr](\xi^{\pm}\cdot\nabla\Phi)^{2}\nonumber\\
	&+2\biggl[2\frac{(\partial_{X}\GG_{4}+2X\partial^{2}_{X}\GG_{4})(\partial_{\Phi}\GG_{4}+2X\partial^{2}_{X\Phi}\GG_{4}-X\partial_{X}\GG_{3})}{1+\GG_{4}-2X\partial_{X}\GG_{4}}\nonumber\\
	&+(\partial_{X}\GG_{3}+X\partial^{2}_{X}\GG_{3}+2X\partial^{3}_{X X\Phi}\GG_{4}+3\partial^{2}_{X\Phi}\GG_{4})\biggr]\xi^{\pm}_{a}\xi^{\pm}_{b}\nabla^{a}\nabla^{b}\Phi.
\end{align}
If $\mathbb{A} \ne 0$ then we must have $\chi=0$, and hence $\lambda=\mu=0$ and $Z_a$ is arbitrary. Hence, in a generic weak-field background, \(\ker \delta P(\xi^{\pm})\) consists of vectors of the form $(t_{ab},0)^T$ where $t_{ab}$ is given by \eqref{t_ans} with $\lambda=\mu=0$. Given that one component of $Z_a$ is ``pure gauge'' (i.e. degenerate with the first term in \eqref{t_ans}), it follows that \(\ker \delta P(\xi^{\pm})\) generically has dimension $7$ and hence the equation of motion is not strongly hyperbolic. 

The only way to escape this conclusion is if the theory is one for which $\mathbb{A}=0$ for {\it any} background. For this to happen, terms with different dependence on the Riemann tensor, $\nabla \Phi$ and $\nabla \nabla \Phi$ have to cancel independently in $\mathbb{A}$. However this cannot happen in the case we are considering. To see this, note that vanishing of the terms of the (schematic) form $R \nabla \Phi \nabla \Phi$ in any background requires $\partial_X^2 \GG_4=0$. But then vanishing of the terms proportional to $R$ requires $\partial_X \GG_4 =0$, contradicting our assumption $\partial_X \GG_4 \ne 0$. Hence in a generic background we have \(\mathbb{A}\neq 0\) and therefore a vector in the kernel must have \(\chi=0\).

In summary, we have shown that when $\partial_X \GG_4 \ne 0$, there does not exist a generalized harmonic gauge for which the equations of motion are strongly hyperbolic in a generic weak-field background. The best one can do is to choose the gauge \eqref{eq:gengauge_4}, for which \(\ker \delta P(\xi^{\pm})\) has dimension $7$ in a generic weak-field background (i.e. $4$ pure gauge elements, and $3$ non-gauge elements). This implies that, in such a background, the matrix $M$ will have two non-trivial Jordan blocks: one in $V^+$ and one in $V^-$. Generically each of these will be $2\times 2$. 

\subsubsection*{Proof of strong hyperbolicity for \(\GG_{4}=\GG_{4}(\Phi)\), $\GG_5=0$}

We continue working with the theory defined by \eqref{quartic}, but now consider the case  $\partial_X \GG_4=0$, i.e., $\GG_{4}=\GG_{4}(\Phi)$.\footnote{An example of such a theory is Brans-Dicke theory \cite{Brans1961} with positive coupling constant $\omega$. After a redefinition of the scalar field, this has  $\GG_{2}=\GG_{3}=0$ and $\GG_{4}=\Phi/\sqrt{2\omega}+\Phi^{2}/(8\omega)$.} We will show that such theories are strongly hyperbolic in a suitable generalized harmonic gauge. The proof is analogous to that for the theory with $\GG_4=\GG_5=0$ so we will be brief. For \(\partial_{X}\GG_{4}=0\), \eqref{eq:2block} reduces to
\begin{align}
	{\rm LHS}=&
	\begin{pmatrix}
		G^{\mu\nu\rho\sigma}[(\GG_{4}-f(f+2))\xi^{\pm}_{\sigma}G_{\rho}{}^{\lambda\alpha\beta}\xi^{\pm}_{\lambda}t_{\alpha\beta}+\xi^{\pm \lambda}\xi^{\pm}_{\sigma}(\mathcal{K}_{\lambda\rho}-\partial_{\Phi}\GG_{4}g_{\lambda\rho})]\chi\\ 
		\\
		G^{\mu\nu\rho\sigma}t_{\mu\nu}\xi^{\pm}_{\rho}\xi^{\pm \lambda}(\mathcal{K}_{\lambda\sigma}-\partial_{\Phi}\GG_{4}g_{\lambda\sigma})+\delta P_{\Phi\Phi}(\xi^{\pm})\chi
	\end{pmatrix}\\
	\nonumber\\
	{\rm RHS}=&
	\begin{pmatrix}
		\xi^{0 \pm}(1+\GG_{4})G^{\mu\nu\rho\sigma}\xi^{\pm}_{\rho}Y^{}_{\sigma}\\
		\\ 
		\xi^{0 \pm}[(\partial_{X} \GG_{3})(\xi^{\pm}\cdot\nabla\Phi)(Y\cdot \nabla\Phi)+\partial_{\Phi}\GG_{4}(\xi^{\pm}\cdot Y)]-\mathcal{K}_{\lambda \sigma}\xi^{\pm \lambda}G^{\mu\nu\sigma 0}\xi^{\pm}_{\mu}Y^{}_{\nu}
	\end{pmatrix} .
\end{align}
Recall that for strong hyperbolicity to hold, this equation must have no solution \((t_{\mu\nu},\chi)^{T}\) when \(Y_{\mu}\neq0\). By the non-degeneracy of \(G^{\mu\nu\rho\sigma}\) we see that if
\begin{equation}
	\GG_{4}-f(f+2)\neq 0
\end{equation}
then we can use the first row of this equation to solve uniquely for $G_{\mu}{}^{\nu\rho\sigma} \xi_\nu^\pm t_{\rho\sigma}$ (the non-transverse part of \(t\)). This can then be substituted into the second row of the equation to give an equation which determines \(\chi\). Hence, if \(\GG_{4}-f(f+2)\neq 0\) then, for any non-zero \(Y_{\mu}\), Eq.~\eqref{eq:2block} has a solution. Therefore for strong hyperbolicity to hold, we need 
\begin{equation}
	\GG_{4}-f(f+2)=0 
	\qquad \Rightarrow \qquad
	f=-1+\sqrt{1+\GG_{4}}
\end{equation}
where we have chosen the root that satisfies the smallness condition \eqref{gauge_small}. With this choice of \(f\), the first row of \eqref{eq:2block} implies
\begin{equation}
\label{Ysol2}
	\xi^{0\pm}Y_{\mu}=\frac{1}{1+\GG_{4}}\xi^{\pm \rho}\tilde{\cal K}_{\rho\mu}\chi.
\end{equation}
where
\begin{equation}
	\tilde{\cal K}_{ab}=\mathcal{K}_{ab}-\partial_{\Phi}\GG_{4}g_{ab}.
\end{equation}
When we plug this into the second row of \eqref{eq:2block} we obtain a linear homogeneous scalar equation for \(\chi\) and \(t_{ab}\). This equation has 11 unknowns and therefore admits a non-trivial solution, generically with $\chi \ne 0$. It follows that if \(Y_{\mu}\) in \eqref{Ysol2} is not vanishing, then strong hyperbolicity fails. This means that strong hyperbolicity requires $\xi^{\pm \rho}\tilde{\cal K}_{\rho\mu}\chi=0$ for arbitrary null \(\xi^{\pm}\). Since generically $\chi \ne 0$, this implies that we must choose our gauge such that \(\tilde{\cal K}_{\mu\nu}=0\). Thus we see that strong hyperbolicity in a generic weak-field background requires us to make the gauge choice
\begin{align}
	\label{eq:gengauge_4_phi}
	f=&-1+\sqrt{1+\GG_{4}}\nonumber \\
	(1+f)\mathcal{H}_{ab}=&\partial_{\Phi}\GG_{4} g_{ab}-\partial_{X}\GG_{3}\nabla_{a}\Phi\nabla_{b}\Phi.
\end{align}
In this gauge, equation \eqref{eq:2block} implies $Y_\mu=0$ so $M$ has no nontrivial Jordan block, i.e., \(M\) is diagonalisable. Note that when $\GG_4=0$ this reduces to the gauge choice \eqref{eq:gengauge_3}.

Diagonalizability of $M$ is a necessary condition for strong hyperbolicity to hold. It ensures the existence of a positive definite symmetrizer $K$ satisfying \eqref{Hcond2}. But we need to check that the remaining conditions in the definition of strong hyperbolicity are satisfied. In particular, we need to prove that $K$ depends smoothly on $\xi_i$. To do this, recall that $K$ is constructed from the matrix $S$ which diagonalizes $M$, as explained above \eqref{Hcond}. $S$ is the matrix whose columns are the eigenvectors of $M$. Hence if the eigenvectors of $M$ depend smoothly on $\xi_i$ then so does $K$. We will explicitly construct the eigenvectors of $M$ to demonstrate that they depend smoothly on $\xi_i$. 

Recall that the eigenvectors of $M$ have the form \eqref{vhorn} where $T$ satisfies \eqref{PT}. In the gauge \eqref{eq:gengauge_4_phi}, we have
\begin{equation}
	\delta P_{gg}(\xi^{\pm})=\delta P_{g\Phi}(\xi^{\pm})=\delta P_{\Phi g}(\xi^{\pm})=0
\end{equation}
which implies that any vector of the form $T=(t_{ab},0)^T$ satisfies \eqref{PT} when $\xi=\xi^\pm$. This proves that the eigenvalues $\xi_0^\pm$ each have degeneracy $10$. If we choose a basis of symmetric tensors $t_{ab}$ that is independent of $\xi_i$ then the $\xi_i$-dependence of these eigenvectors arises only through the $\xi_0$ in \eqref{vhorn}, which implies that these $20$ eigenvectors depend smoothly on $\xi_i$. A calculation reveals that the final two eigenvectors have $T=(t_{ab},1)^T$ where
\be
\label{eq:scalar_eigenvector}
 t_{ab} = -\frac{\partial_{X}\GG_{3}}{1+\GG_{4}} \left[ \nabla_a \Phi \nabla_b \Phi + g_{ab} X \right]-\partial_{\Phi}\log(1+\GG_{4})g_{ab}
\ee
and eigenvalues $\xi_0$ determined by 
\be
\label{fdef}
0 = f^{\mu\nu}\xi_\mu \xi_\nu  \equiv -P_{\Phi\Phi}(\xi) - \frac{1}{(1+\GG_{4})}\left[X^{2}(\partial_{X}\GG_{3})^{2}+2(\partial_{\Phi}\GG_{4})^{2}\right]\xi^{2} .
\ee
For a weak field background, $f^{\mu\nu}$ is close to $g^{\mu\nu}$ and is therefore a Lorentzian metric with $f^{00} \ne 0$. This ensures that there will be two real eigenvalues $\xi_0$ depending smoothly on $\xi_i$. As before, the eigenvectors depend on $\xi_i$ only through $\xi_0$ and are therefore smooth. Hence all eigenvectors have the required smoothness in $\xi_i$ so the symmetrizer is smooth. This establishes strong hyperbolicity in the gauge \eqref{eq:gengauge_4_phi}.\footnote{\label{rendall} Actually we should also check the inequality below \eqref{Hcond2}. This follows trivially if we restrict to a compact region of spacetime. For the \(\LL_{1}+\LL_{2}\) theory, a stronger result can be obtained \cite{Rendall2006}: this theory is {\it symmetric} hyperbolic even outside of the ``weak field'' regime provided that $1+\partial_{X}\GG_{2}>0$ and  $1+\partial_{X}\GG_{2}+2X(\partial^{2}_{X}\GG_{2}) >0$. In our case, the smallness condition \eqref{eq:horndeski_smallness_2} implies that these conditions are satisfied.} 

\subsubsection*{Failure of strong hyperbolicity if $\GG_5 \ne 0$}

Finally, we include the term \(\LL_{5}\) into the Lagrangian. We refer to Ref. \cite{Kobayashi2011} for the explicit form of the equations of motion. With \(\GG_{5} \ne 0\) we expect to encounter similar issues as those we encountered in theories with $\partial_X \GG_4 \ne 0$, $\GG_5=0$. This can be seen easily if we consider the case $\GG_4=0$ with $\partial_X \GG_5 =0$, i.e., 
\begin{equation}
	\GG_{5}=\GG_{5}(\Phi).
\end{equation}
In this case we can write \cite{Kobayashi2011,Bettoni2013} 
\begin{align}
	\LL_{5}&=\GG_{5}(\Phi)G_{ab}\nabla^{a}\nabla^{b}\Phi\nonumber\\
	& =-\partial_{\Phi}\GG_{5} X R -\partial_{\Phi}\GG_{5}\delta^{ac}_{bd}\nabla_{a}\nabla^{b}\Phi\nabla_{c}\nabla^{d}\Phi+3\partial^{2}_{\Phi}\GG_{5} X \square \Phi - 2 \partial^{3}_{\Phi}\GG_{5} X^{2}+\ldots
\end{align}
where the ellipsis denotes a total derivative term which does not contribute to the equations of motion. Therefore we can rewrite \(\LL_{5}\) as a sum of lower order Lagrangians
\begin{equation}
	\LL_{5}=\tilde{\LL}_{4}+\tilde{\LL}_{3}+\tilde{\LL}_{2}
\end{equation}
where
\begin{align}
	\tilde{\GG}_{4}=-\partial_{\Phi}\GG_{5}X\quad \tilde{\GG}_{3}=3\partial^{2}_{\Phi}\GG_{5}X \quad \tilde{\GG}_{2}=-2\partial^{3}_{\Phi}\GG_{5}X^{2}.
\end{align}
Since $\partial_{X}\tilde{\GG}_{4}\neq 0$, our previous results imply that there is no generalized harmonic gauge for which theory is strongly hyperbolic in a generic weak-field background.\footnote{Note that if $\partial_\Phi \GG_5=0$ then $\GG_5$ is a constant, which implies that $\LL_5$ is a total derivative.} Of course, we could cure the above problem by adding a $\GG_4$ term to cancel $\tilde{\GG}_4$ but then all we are doing is reducing the theory to a theory with $\GG_4 = \GG_5=0$. The issue is that there is degeneracy between the $\LL_5$ term and the other terms in the Lagrangian. We can remove this degeneracy by supplementing the conditions \eqref{horn_cond} with
\be
\label{GG5cond}
 \GG_5 (\Phi,0)=0
\ee
which means that non-trivial $\GG_5$ must depend on $X$. 

Although we have not analyzed it in detail, there seems very little chance that a theory with $\partial_X \GG_5  \ne 0$ could be strongly hyperbolic in some generalized harmonic gauge for a generic weak-field background. Indeed, we expect such a theory to exhibit even worse behaviour than the $\partial_X \GG_4 \ne 0$, $\GG_5=0$ case in the following sense. We mentioned above, that for the latter theory, one can find a generalized harmonic gauge for which $M$ generically has just $2$ non-trivial Jordan blocks. We expect a $\partial_X \GG_5 \ne 0$ theory to be worse in the sense that, generically, for any generalized harmonic gauge, $M$ will have $8$ non-trivial ($2\times 2$) Jordan blocks, $4$ in each of $V^+$ and $V^-$. In other words, for this theory, all pure gauge eigenvectors will be associated to non-trivial Jordan blocks, just as for (harmonic gauge) Lovelock theory. This is consistent with the fact that some theories of this type can be obtained by dimensional reduction of Lovelock theories.

\subsubsection*{Summary of results for linearized theory}

We have proved that, if $\partial_X \GG_4 \ne 0$ and $\GG_5 = 0$ then there exists no generalized harmonic gauge for which linearized Horndeski theory is strongly hyperbolic in a generic weak-field background. However, if $\partial_X \GG_4 = \GG_5 =0$ then there exists a unique generalized harmonic gauge for which linearized Horndeski theory is strongly hyperbolic in a generic weak-field background. We have not analyzed the case $\GG_5 \ne 0$ in detail but we believe that, once degeneracy with other terms has been eliminated via \eqref{GG5cond}, this case is not compatible with strong hyperbolicity in a generic weak-field background either. 

This means that any Horndeski theory (satisfying \eqref{horn_cond}) for which there exists a generalized harmonic gauge such that the linearized equation of motion is strongly hyperbolic around a generic weak-field background can be obtained from a Lagrangian of the form 
\begin{equation}
\label{hyp_horn}
	\LL=R+X-V(\Phi)+\GG_{2}(\Phi,X)+\GG_{3}(\Phi,X)\square \Phi+\GG_{4}(\Phi)R.
\end{equation}
More general Horndeski theories will fail to be strongly hyperbolic around a generic weak-field background in any generalized harmonic gauge. 

Causal properties of theories of the form \eqref{hyp_horn} have been discussed in Ref.~\cite{Deffayet2010}.\footnote{\label{conf_trans} Ref. \cite{Deffayet2010} assumed $\GG_4 =0$ but for a theory of the form \eqref{hyp_horn} we can always set $\GG_4=0$ using a field redefinition, specifically a conformal transformation.} It is interesting to discuss causality using our results above. We showed above that, in an appropriate generalized harmonic gauge, a null co-vector $\xi_a$ is characteristic if, and only if, either $g^{ab} \xi_a \xi_b =0$ or $f^{ab} \xi_a \xi_b =0$, where $f^{ab}$ is defined by \eqref{fdef}. 
Furthermore, if $\xi_a$ satisfies the former condition then $P(\xi)$ generically has a $10$-dimensional kernel consisting of vectors of the form $(t_{ab},0)$ for general $t_{ab}$, whereas if $\xi_a$ satisfies the latter condition then $P(\xi)$ generically has a $1$-dimensional kernel consisting of vectors of the form $(t_{ab},1)$ with $t_{ab}$ given by \eqref{eq:scalar_eigenvector}. 
Hence, roughly speaking, causality for the $10$ tensor degrees of freedom is determined by $g_{ab}$ whereas causality for the $1$ scalar degree of freedom is determined by $f_{ab}$, the inverse of $f^{ab}$. This agrees with Ref.~\cite{Deffayet2010}. Of course these degrees of freedom are coupled together so causality for the theory as a whole is determined by both metrics $g_{ab}$ and $f_{ab}$. 
More precisely, the characteristic surfaces of the theory are surfaces which are null w.r.t. either $g_{ab}$ or $f_{ab}$. 

\subsubsection*{Nonlinear considerations}

The above discussion shows that there exists a preferred generalized harmonic gauge \eqref{eq:gengauge_4_phi} for which a theory of the form \eqref{hyp_horn} is strongly hyperbolic when linearized around a generic weak field background. We can now ask: does this generalized harmonic gauge condition for the linearized theory arise by linearizing a generalized harmonic gauge condition for the nonlinear theory?

Consider a nonlinear generalized harmonic gauge condition of the form
\be
 \frac{1}{\sqrt{-g}} \partial_\nu \left( \sqrt{-g} g^{\mu\nu} \right) = J^\mu(g,\Phi,\partial \Phi).
\ee
Note that we would not want $J^\mu$ to depend on second or higher derivatives of $\Phi$ because this would give a gauge-fixed equation of motion involving third derivatives of $\Phi$. 

Linearizing around a general background solution gives
\be
 \nabla_\nu h^{\mu\nu} - \frac{1}{2} \nabla^\mu h^\nu_\nu + \frac{\partial J^\mu}{\partial (\partial_\nu \Phi)} \partial_\nu \psi = \ldots
\ee
where the ellipsis denotes terms that do not involve derivatives of $h_{\mu\nu}$ or $\psi$ and therefore do not influence hyperbolicity. Comparing with \eqref{eq:ghg} we see that the linearized gauge condition has
\be
\label{HJeq}
 \frac{{\cal H}^{\mu\nu}}{1+f}  = -  \frac{\partial J^\mu}{\partial (\partial_\nu \Phi)}.
\ee
It follows that the functions appearing in the linearized gauge condition must satisfy the integrability condition
\be
 \frac{\partial}{\partial (\partial_\rho \Phi)} \left( \frac{{\cal H}^{\mu\nu}}{1+f} \right)=  \frac{\partial}{\partial (\partial_\nu \Phi)} \left( \frac{{\cal H}^{\mu\rho}}{1+f} \right).
\ee
Plugging in the functions \eqref{eq:gengauge_4_phi}, this equation reduces to
\be
\partial_X \GG_3 \left( g^{\mu\nu} \partial^\rho \Phi - g^{\mu\rho} \partial^\nu \Phi \right) = 0.
\ee 
By contracting this equation it is easy to see that the only way this can hold in a generic background is if $\partial_X \GG_3 =0$. But then $\GG_3$ is independent of $X$, so \eqref{horn_cond} implies $\GG_3 = 0$.\footnote{If $\GG_3$ is independent of $X$ then a term in the action of the form $\LL_3$ is degenerate with a term of the form $\LL_2$ and the conditions \eqref{horn_cond} were imposed to eliminate this degeneracy.} If $\GG_3=0$ then we can find a source function $J^\mu$ consistent with equation \eqref{HJeq}:
\be
\label{Jchoice}
 J^\mu = -\frac{\partial_\Phi \GG_4}{1+\GG_4} \partial^\mu \Phi.
\ee
In summary, we have imposed the requirement that the preferred generalized harmonic gauge condition for the linearized theory arises by linearizing a generalized harmonic gauge condition for the nonlinear theory. The result is that this requirement excludes theories with non-trivial $\GG_3$. So demanding that there exists a generalized harmonic gauge for which the nonlinear theory is strongly hyperbolic in a generic weak-field background restricts the theory to one of the form
\be
\label{hyp_horn_nonlinear}
\LL=R+X-V(\Phi)+\GG_{2}(\Phi,X)+\GG_{4}(\Phi)R.
\ee 
Since $\GG_4$ can be eliminated by a field redefinition (footnote \ref{conf_trans}), this theory is equivalent to Einstein gravity coupled to a ``k-essence'' theory. With the gauge choice \eqref{Jchoice}, this theory is not just strongly hyperbolic, it is {\it symmetric} hyperbolic (see footnote \ref{rendall}). 

\section{Discussion}

\label{sec:discuss}

We have shown that, in harmonic gauge, the linearized equation of motion of a Lovelock theory is always weakly hyperbolic in a weakly curved background. However, it is not strongly hyperbolic in a generic weak-field background. We have shown that, in a generalized harmonic gauge, the linearized equation of motion of a Horndeski theory is always weakly hyperbolic in a weak-field background. For some Horndeski theories, a generalized harmonic gauge can be found for which the linearized equation of motion is also strongly hyperbolic in a weak field background. In particular this is true for theories of the form \eqref{hyp_horn}. However, for more general Horndeski theories, we have shown that there is no generalized harmonic gauge for which the equation of motion is strongly hyperbolic in a generic weak-field background. Furthermore, even for theories of the form \eqref{hyp_horn}, imposing the requirement that the gauge condition for the linearized theory is the linearization of a generalized harmonic gauge condition for the nonlinear theory restricts the theory further, to one of the form \eqref{hyp_horn_nonlinear}. 

Without strong hyperbolicity, the best one can hope for is that the linearized equation of motion is locally well-posed with a ``loss of derivatives''. This means that the $k$th Sobolev norm $H^{k}$ of the fields at time $t$ cannot be bounded in terms of its initial value but only in terms of the initial value of some higher Sobolev norm $H^{k+l}$ with $l>0$. Whether or not even this can be done depends on the nature of the terms with fewer than two derivatives  in the equation of motion \cite{Kreiss1989}. But even if this can be achieved, the loss of derivatives is likely to be fatal for any attempt to prove that the {\it nonlinear} equation is locally well-posed in some Sobolev space, as is the case for the Einstein equation.\footnote{\label{gevrey} It is conceivable that one might have local well-posedness in some much more restricted function space, such as a Gevrey space.} This is because establishing well-posedness for a nonlinear equation usually involves a ``bootstrap'' argument in which one assumes some bound on the $H^{k}$ norm and then uses the energy estimate to improve this bound, thereby closing the bootstrap. This is not possible if the energy estimate exhibits a loss of derivatives. 

Note that our result is a statement about the {\it full} equations of motion. If one restricts the equations of motion by imposing some symmetry on the solution (e.g. spherical symmetry) then it is possible that the resulting equations might be strongly hyperbolic. This is because the resulting class of background spacetimes would be non-generic and, as we have seen, for non-generic backgrounds it is possible for the equation of motion to be strongly hyperbolic even if it is not strongly hyperbolic for a generic background. 

Our results demonstrate that we do not have local well-posedness for the harmonic gauge Lovelock equation of motion for {\it general} initial data. So the situation is worse than for the Einstein equation, for which the harmonic gauge equation of motion is locally well-posed for any initial data \cite{Choquet-Bruhat1952}. But in practice we are not interested in general initial data, but only in initial data satisfying the harmonic gauge condition. Since the failure of strong hyperbolicity appears to be associated to modes which violate the harmonic gauge condition, perhaps we could restrict to initial data satisfying this condition {\it exactly} and thereby obtain a well-posed problem. One could not do this numerically on a computer because the gauge condition could never be imposed exactly -- there would always be numerical error. But perhaps this could be done in principle. One way to proceed would be to consider sequences of analytic initial data, satisfying the gauge condition, which approach some specified smooth initial data. For analytic data one can solve the equation of motion locally \cite{Choquet-Bruhat1988}. If one could prove that the resulting analytic solution satisfies an energy estimate without a loss of derivatives (because it satisfies the gauge condition), then perhaps it would be possible to establish local well-posedness. Having said this, we note that one could make exactly the same remarks about the Einstein equation written in a ``bad'' (non-strongly hyperbolic) gauge so it is far from clear that this method has any chance of succeeding.

If the equation of motion is not strongly hyperbolic in (generalized) harmonic gauge then could there be some other gauge in which it is strongly hyperbolic? For example, maybe one could modify the (generalized) harmonic gauge condition to include additional terms involving first derivatives of $h_{ab}$, contracted in some way with the background curvature tensor (or scalar field). But this raises the question of whether it is always possible to impose the new gauge condition via a gauge transformation. This would involve solving an equation for the gauge parameters. We would then have to analyze whether this new equation has a well-posed initial value problem, and whether the resulting gauge condition is propagated by the gauge-fixed equation of motion. This may amount to analyzing equations that suffer from the same kind of problems as the equations we have discussed in this paper. 

In this paper, we have been working with equations of motion for the metric. An alternative approach would be to derive an equation of motion for curvature. The Bianchi identity can be used to write $\nabla^e \nabla_e R_{abcd}$ in terms of second derivatives of the Ricci tensor, and terms with fewer than two derivatives of curvature. For the Einstein equation, one can eliminate the Ricci tensor to obtain a nonlinear wave equation for the Weyl tensor. This equation is strongly hyperbolic and admits a well-posed initial value problem. For a Lovelock theory one cannot solve explicitly for the Ricci tensor but one could still replace the Ricci tensor terms using the expression obtained from the equation of motion of the theory. This gives an equation of motion for the Riemann tensor. In contrast with what happens for the Einstein equation, the resulting equation is subject to a constraint, which is simply the Lovelock equation of motion. If this constraint is satisfied by the initial data then it will be satisfied by any solution of the equation of motion for the Riemann tensor. The situation looks analogous to the case of the harmonic gauge equation of motion for the metric, but with more indices. It seems very likely that this equation of motion for the Riemann tensor will fail to be strongly hyperbolic in a generic background. 

Another approach would be to investigate equations of motion based on a space-time decomposition of the metric, as in the ADM formalism. It is known that the ADM formulation of the Einstein equation gives equations that are not strongly hyperbolic \cite{Nagy2004}. However, suitable modification of the ADM method gives equations that {\it are} strongly hyperbolic \cite{Nagy2004,Rodnianski2014a}. Perhaps something similar would work for Lovelock or Horndeski theories. However, it appears that there is no obvious way of extending the approaches used for the Einstein equation to Lovelock theories \cite{KeirUnpublished}.

Of course there is also the possibility that these theories do not admit a locally well-posed initial value problem, or that one only has well-posedness for some highly restricted space of initial data. This would lead to the satisfying conclusion that these modifications of the Einstein equation can be shown to be unviable as physical theories solely on the basis of the classical initial value problem for weak fields. 

\subsection*{Acknowledgments}

We are grateful to M. Dafermos, J. Keir, J. Luk, J. Santos, A. Vikman and H. Witek for useful conversations. GP is supported by an STFC studentship.

\bibliographystyle{JHEP}
\bibliography{hyperbolicity_lovelock_horndeski}

\end{document}